\documentclass{article}

\PassOptionsToPackage{numbers, compress}{natbib}

\usepackage[main, final]{neurips_2026}

\usepackage[utf8]{inputenc}
\usepackage[T1]{fontenc}
\usepackage{hyperref}
\hypersetup{pdftitle={UMAS: System-Level Uncertainty Quantification for Multi-Agent LLM Systems}, pdfauthor={Hanwen Li, Jinhao Duan, Xiaoshuang Shi, Yue Zhang, Tianlong Chen, Kaidi Xu, Chenxi Yuan}}
\usepackage{url}
\usepackage{booktabs}
\usepackage{microtype}
\usepackage{xcolor}
\usepackage{graphicx}
\usepackage{amsmath}
\usepackage{amssymb}
\usepackage{multirow}
\usepackage{pifont}
\usepackage{algorithm}
\usepackage{algpseudocode}
\usepackage{subcaption}
\usepackage{tcolorbox}
\usepackage{xltabular}

\def\gE{{\mathcal{E}}}
\def\gG{{\mathcal{G}}}
\def\gV{{\mathcal{V}}}
\def\gY{{\mathcal{Y}}}

\def\sP{{\mathbb{P}}}
\def\sR{{\mathbb{R}}}

\newcommand{\E}{\mathbb{E}}

\DeclareMathOperator*{\argmin}{arg\,min}

\newcommand{\methodname}{\ensuremath{\mathrm{UMAS}}}
\newcommand{\methodfix}{\ensuremath{\mathrm{UMAS}_{\mathit{FIX}}}}

\newcommand{\masonename}{Debate}

\newcommand{\mastwoname}{Dynamic LLM-Agent Network}
\newcommand{\mastwoabb}{DyLAN}

\definecolor{reference}{RGB}{0, 143, 19}
\definecolor{query}{RGB}{66, 111, 245}

\title{\methodname: System-Level Uncertainty Quantification for Multi-Agent LLM Systems}

\author{%
  Hanwen Li\textsuperscript{1} \quad Jinhao Duan\textsuperscript{2}\thanks{Corresponding authors.} \quad Xiaoshuang Shi\textsuperscript{3} \quad Yue Zhang\textsuperscript{4} \\
  \textbf{Tianlong Chen\textsuperscript{2} \quad Kaidi Xu\textsuperscript{5} \quad Chenxi Yuan\textsuperscript{1}\footnotemark[1]} \\[0.4em]
  \textsuperscript{1}New Jersey Institute of Technology \quad
  \textsuperscript{2}University of North Carolina at Chapel Hill \\
  \textsuperscript{3}University of Electronic Science and Technology of China \quad
  \textsuperscript{4}Shandong University \\
  \textsuperscript{5}City University of Hong Kong \\
  \small\texttt{hl544@njit.edu}, \texttt{jinhao@cs.unc.edu}, \texttt{xsshi2013@gmail.com}, \texttt{zyueinfosec@sdu.edu.cn}, \\
  \small\texttt{tianlong@cs.unc.edu}, \texttt{kaidixu@cityu.edu.hk}, \texttt{chenxi.yuan@njit.edu}
}

\begin{document}

\maketitle

\begin{abstract}
  Multi-Agent Systems (MAS) are a promising paradigm for improving the reasoning capabilities of large language models, yet they remain prone to hallucinations and erroneous outputs. Uncertainty Quantification (UQ) is crucial for assessing system reliability, but existing methods are largely designed for single-agent settings; how to exploit MAS structure for UQ while keeping inference cost-neutral remains underexplored.
We propose \methodname, a parameter-free, cost-neutral, system-level UQ framework that estimates the credibility of inter-agent influence from each generation's intrinsic confidence. \methodname~uses these credibility signals to modulate how uncertainty is updated at each node, and then aggregates node-level uncertainty into a system-level estimate. Extensive evaluations on Debate and DyLAN across diverse benchmarks show that \methodname~consistently outperforms state-of-the-art baselines by an average of 10.54 AUROC points. Beyond uncertainty estimation, \methodname~enables hallucination detection and uncertainty-aware answer selection, improving MAS accuracy by up to 12.6 points on specific tasks while enhancing reliability. Code is available at \url{https://github.com/Dby52-04/UMAS}.
\end{abstract}

\section{Introduction}\label{sec:introduction}

Large Language Models (LLMs) have achieved strong performance across many applications~\cite{zhao2023survey}, yet they still struggle with factuality and complex reasoning, leading to hallucinations and errors~\cite{huang2025survey, lin-etal-2022-truthfulqa, ji2023survey}. Prior work mitigates these issues via Chain-of-Thought~\cite{wei2022chain}, self-consistency~\cite{wang2022self}, and Retrieval-Augmented Generation~\cite{gao2023retrieval, ng2025rag}, and multi-agent systems (MAS) further extend capability through collaborative reasoning~\cite{li2024survey, chen2024survey}. However, hallucinations persist even in MAS~\cite{ye2025maslab}, making reliable detection of erroneous agent outputs essential for trustworthy deployment.

\begin{figure}[!thp]
    \centering
    \includegraphics[width=1\textwidth]{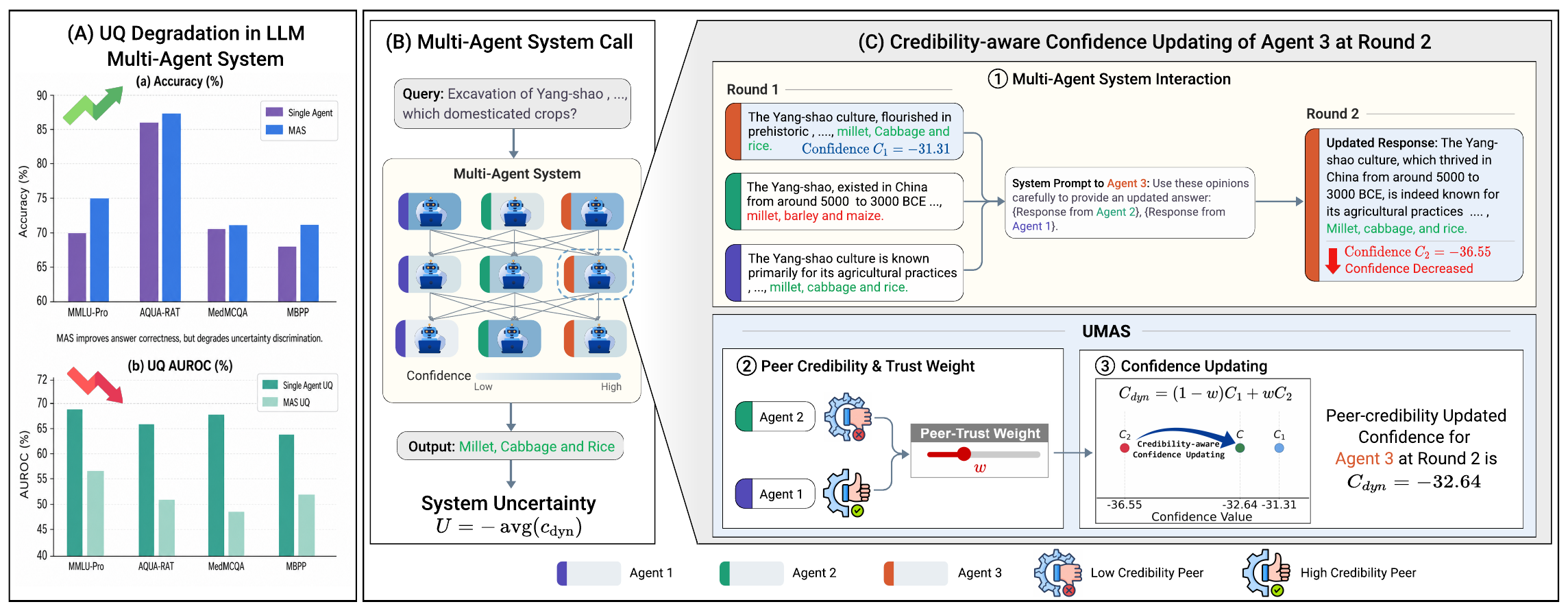}
    \caption{\textbf{Left (A):} Applying standard single-agent UQ to the final MAS output exhibits degraded reliability (lower AUROC) despite improved task accuracy. \textbf{Right (B \& C):} Overview of the \methodname~framework during a collaborative update. A focal agent (e.g., Agent 3) revises its response using peer generations. \methodname~estimates node-level uncertainty and dynamically calibrates it based on peer credibility. Peer credibility (denoted by the gear icons) is evaluated to compute a trust weight, generating a credibility-aware updated confidence score ($c_{\mathrm{dyn}}$).}
    \label{fig:illustration}
\end{figure}

Uncertainty Quantification (UQ) is a key tool for evaluating the reliability and trustworthiness of LLM generations. As illustrated in Figure~\ref{fig:illustration}, an important issue emerges: although MAS can achieve higher task accuracy than a single LLM (Appendix~\ref{app:mas_accuracy}), standard UQ applied to MAS output can become substantially less reliable. We identify two major limitations of existing UQ methods when applied to MAS.
(i) \textbf{Structure misalignment:} Most UQ methods estimate uncertainty solely from the final response to a query~\cite{kuhn2023semantic,vashurin2025uncertaintyquantificationllmsminimum, du2024haloscope}. However, an MAS output is produced through a multi-round, multi-agent interaction trajectory; restricting UQ to the terminal answer ignores this structure and can miss how uncertainty is propagated and amplified through inter-agent influence.
(ii) \textbf{Efficiency:} MAS also substantially increases inference cost. As shown in Table~\ref{tab:long_example}, while MAS architectures improve response accuracy, they incur significant overhead. Meanwhile, many UQ approaches~\cite{kuhn2023semantic, duan2024shifting, vashurin2025uncertaintyquantificationllmsminimum, lin2023generating} require multiple samples per query, which becomes prohibitively expensive for MAS due to the length and multiplicity of multi-agent generations.

These limitations motivate our primary research question: can we develop an MAS-specific UQ framework that satisfies two core criteria? (i) \textbf{Structure-aware:} leverage the structure and interaction trajectory of MAS, rather than treating uncertainty solely as a property of the final system output. (ii) \textbf{Cost-neutral:} maintain near-zero additional inference overhead, requiring no extra sampling or additional model invocations.

To answer this question, we propose \methodname, a system-level UQ framework for MAS. Instead of computing uncertainty only on the terminal response, \methodname~assigns node-level uncertainty to each agent response at each interaction round and derives a system-level uncertainty by aggregating these node-level estimates over the MAS interaction graph. The key design question is how to assign node-level uncertainty in light of inter-agent interactions: in collaborative MAS protocols, an agent revises its response by incorporating peer generations, but blindly trusting peer information can distort uncertainty---analogous to anchoring on an unreliable peer's claim. \methodname~therefore estimates the credibility of each inter-agent influence from the generations' intrinsic confidence and uses these signals to modulate how uncertainty is updated during interaction. Figure~\ref{fig:illustration} illustrates the procedure for a particular update instance. This credibility-aware UQ prevents unreliable peer evidence from inducing overconfident or spurious uncertainty.

We evaluate \methodname~under two representative collaborative MAS paradigms: \masonename~\cite{du2023improving}, a canonical collaboration-based protocol, and \mastwoname~(\mastwoabb)~\cite{liu2024dynamic}, an adaptive framework that supports iterative team reconfiguration during execution. We instantiate both architectures with two widely used LLM backbones, GPT-4o mini~\cite{achiam2023gpt} and Llama 4 Scout~\cite{meta2025llama4}, and benchmark on five diverse tasks: MMLU-Pro (general knowledge)~\cite{wang2024mmlu}, AQUA-RAT (mathematics)~\cite{ling2017program}, MedMCQA (medicine)~\cite{pal2022medmcqa}, MBPP (programming)~\cite{austin2021program}, and GSM-Hard (mathematics)~\cite{gao2022pal}.
Across architectures, models, and datasets, \methodname~improves the reliability of uncertainty estimates over strong UQ baselines~\cite{kuhn2023semantic, lin2023generating, duan2024shifting, aichberger2024rethinking, vashurin2025uncertaintyquantificationllmsminimum}. Compared to the best-performing baseline in each setting, \methodname~achieves an average AUROC gain of 10.54 points across both backbones. Finally, we show that the resulting uncertainty improvements translate into downstream gains for selective prediction and hallucination detection in collaborative MAS, improving overall system reliability and enabling effective detection of erroneous responses. Our contributions are summarized as follows:
\begin{itemize}
    \item We identify a failure mode that harms UQ reliability in MAS: although MAS often improves task accuracy, applying standard UQ can significantly degrade uncertainty reliability. We attribute this to a structural mismatch: MAS uncertainty depends on the multi-round interaction and the credibility of inter-agent influence.

    \item We propose \methodname, a parameter-free and cost-neutral UQ framework that assigns node-level uncertainty to each agent response at each interaction round and aggregates it into a system-level uncertainty.

    \item We evaluate \methodname~across two collaborative MAS paradigms, two LLM backbones, and five benchmarks. We further show that these UQ improvements translate into better selective prediction and hallucination detection in MAS, thereby enhancing the overall performance and reliability of the MAS.
\end{itemize}

\section{Related Works}

\paragraph{Uncertainty Quantification.}\label{para:uncertainty_estimation} Uncertainty Quantification (UQ) is critical for assessing LLM reliability, with existing literature largely focused on single-turn tasks~\cite{shelmanovvashurin2025}. Methodologies generally fall into two categories: (i) information-theoretic approaches, which efficiently derive uncertainty from token probabilities but often overlook semantic variability, and (ii) consistency-based strategies, such as Semantic Entropy~\cite{kuhn2023semantic}, which utilize multi-sampling to analyze semantic alignment across generations. Recent hybrid frameworks like CoCoA~\cite{vashurin2025uncertaintyquantificationllmsminimum} aggregate confidence and consistency to bridge these perspectives. Furthermore, UQ has expanded into multi-step decision-making; methods like SAUP~\cite{zhao-etal-2025-uncertainty}, for instance, train a Hidden Markov Model to assign a situational weight to the uncertainty of each step. However, this method requires ground-truth labels from the test domain and is not designed for MAS. UProp~\cite{duan2025uprop} samples multiple trajectories of a single agent's decision chain, a special case of our trajectory graph; \methodname~scores one realized multi-agent trajectory.

\paragraph{Multi-Agent Systems.} To address the limitations of single-agent models, Multi-Agent Systems (MAS) employ collaborative, multi-turn interactions~\cite{ye2025maslab}. Foundational frameworks like CAMEL~\cite{li2023camel} and AutoGen~\cite{wu2024autogen} utilize role-playing mechanisms to facilitate task completion, while debate-oriented architectures~\cite{du2023improving} leverage dialectical strategies for solution refinement. Advancing beyond static topologies, DyLAN~\cite{liu2024dynamic} introduces dynamic team optimization, and MAS-GPT~\cite{ye2025mas} generates query-specific architectures. Concurrently, approaches like DiverseAgentEntropy~\cite{feng2025rethinking} integrate uncertainty metrics to prioritize reliable agents. Our work focuses on the collaborative update mechanism, a fundamental component across these paradigms, to enhance UQ within MAS.

\section{Preliminary}

\paragraph{Confidence Estimation.}
LLMs are autoregressive generative models. Letting $p_\theta$ denote the parameterized LLM, the probability of output $y$ given input $x$, referred to as the Maximum Sequence Probability (MSP), factorizes over tokens; we work with its logarithm throughout:
\begin{equation}
    \log P(y \mid x) = \sum_{t=1}^{T} \log p_\theta(y_t \mid y_{<t}, x),
    \label{eq:msp}
\end{equation}
where $y = (y_1, \dots, y_T)$ is the output sequence of length $T$ and $y_{<t}$ is its prefix. Its negation $-\log P(y \mid x)$ is the standard MSP uncertainty score~\cite{shelmanovvashurin2025}. MSP serves as a computationally efficient confidence proxy~\cite{kadavath2022language}.
\paragraph{Multi-Agent System.} Following the framework proposed in~\cite{qian2024scaling}, we formulate the MAS as a directed graph architecture $\gG=(\gV, \gE)$. We define $N$ as the number of agents and $M$ as the total number of communication rounds (equivalent to layers in the graph). The vertex set $\gV$ represents the complete history of agent states:
\begin{equation*}
    \gV = \left\{ v_i^{(t)} \mid i \in \{1, \dots, N\}, \, t \in \{1, \dots, M\} \right\},
\end{equation*}
where $v_i^{(t)}$ denotes the state of the $i$-th agent at round $t$. Let $g(v_i^{(t)})$ represent the content generated by the LLM for this node. The edge set $\gE$ is constructed based on the information dependency flow. Specifically, if the generation of $v_i^{(t)}$ depends on aggregating a set of predecessor states from the previous round, denoted as $\sP_i^{(t)} \subseteq \{v_1^{(t-1)}, \dots, v_N^{(t-1)}\}$, we define the edges as:
\begin{equation*}
    \gE = \bigcup_{t}^{M} \bigcup_{i}^{N} \left\{ \big\langle u, v_i^{(t)} \big\rangle \mid u \in \sP_i^{(t)} \right\}.
\end{equation*}

\section{Method}\label{sec:method}

\paragraph{Node-level Confidence Score.} While various uncertainty estimation methods exist, such as Semantic Entropy (SE)~\cite{kuhn2023semantic} and Predictive Entropy (PE)~\cite{fomicheva2020unsupervised}, we use Maximum Sequence Probability (MSP; Eq.~\ref{eq:msp}) because it is cost-neutral for a completed generation. Unlike sampling-based methods, MSP estimates confidence directly from generation logits and requires no additional inference passes. Formally, let $x_i^{(t)}$ denote the input context for node $v_i^{(t)}$. We assign the node a scalar confidence score $c_i^{(t)}$, defined as the log-probability of the generated response conditioned on the input:
\begin{equation}
    c_i^{(t)} = \log P\left(g(v_i^{(t)}) \mid x_i^{(t)}\right),
\end{equation}
where $\log P$ is the log-MSP defined in Eq.~\ref{eq:msp}. Hence $c_i^{(t)}<0$, and a larger (less negative) value indicates higher confidence.

\methodname~extends this node-level confidence to the MAS setting by tracking how confidence evolves through inter-agent interactions. In addition to the raw confidence $c_i^{(t)}$, each node maintains a dynamic confidence $c_{i,\mathrm{dyn}}^{(t)}$, which adjusts the node's confidence according to the credibility of the peer evidence used to produce it.

\subsection{Peer Credibility Estimation}
In a collaborative MAS, an agent often revises its response by conditioning on generations from other agents. The reliability of these peer references should therefore affect the confidence assigned to the revised response. For a node $v_i^{(t)}$, we denote its peer reference set as $\sR_i^{(t)}=\{g(u)\mid \langle u, v_i^{(t)}\rangle \in \mathcal{E}\}$.

Rather than treating all references equally, \methodname~estimates a credibility score for each directed update edge. For an edge $\langle v_j^{(t-1)}, v_i^{(t)} \rangle \in \mathcal{E}$, where peer node $v_j^{(t-1)}$ contributes to the update of $v_i^{(t)}$, we define the raw edge credibility as the peer's relative dynamic confidence with respect to the updater's previous state:
\begin{equation}
    w_{j \to i}^{(t-1)} = \frac{c_{j,\mathrm{dyn}}^{(t-1)} - c_{i,\mathrm{dyn}}^{(t-1)}}{c_{i,\mathrm{dyn}}^{(t-1)}},
    \label{eq:confidence}
\end{equation}
where $c_{j,\mathrm{dyn}}^{(t-1)}$ and $c_{i,\mathrm{dyn}}^{(t-1)}$ are the dynamic confidences from the previous round. The relative form in Eq.~\ref{eq:confidence} makes the score comparable across prompts, datasets, and models. Since $c_{i,\mathrm{dyn}}^{(t-1)}<0$, a peer more confident than the updater's prior state yields a negative score, whereas a less confident, and hence less credible, peer yields a positive score. This gives \methodname~a simple heuristic signal, not a truth detector (Appendix~\ref{app:limitations}), for deciding how strongly peer-supported revisions should affect confidence.

\subsection{Credibility-aware Confidence Updating}
Given the raw edge credibility scores above, we next convert peer evidence into a bounded trust weight and use it to update the agent's dynamic confidence.
\paragraph{Peer-Trust Weighting.}
For each update node $v_i^{(t)}$, \methodname~summarizes all incoming peer evidence into a single node-level trust weight. We first map each raw edge score $w_{j\to i}^{(t-1)}$ to $(0,1)$ with a sigmoid:
\begin{equation}
    \hat{w}_{j\to i}^{(t-1)} = \sigma\!\left(w_{j\to i}^{(t-1)}\right)\in(0,1).
    \label{eq:normalize_weight}
\end{equation}
where $\sigma(x)=\frac{1}{1+\exp(-x)}$. This mapping stabilizes the update by preventing extreme relative-confidence ratios from producing unbounded influence. It also preserves the ordering of the credibility signal: equal confidence maps to $0.5$, more credible peers map below $0.5$, and less credible peers map above $0.5$. We then average the normalized incoming edge scores and take the complement to obtain the node-level trust weight, so that more credible peers yield a larger weight:
\begin{equation}
    w_i^{(t)}=1-\E_{j:\langle v_j^{(t-1)},v_i^{(t)}\rangle\in\mathcal{E}} \left[\hat{w}_{j\to i}^{(t-1)}\right].
    \label{eq:update_weight}
\end{equation}
The weight in Eq.~\ref{eq:update_weight} should be interpreted as the effective credibility of the current peer-supported update. Appendix~\ref{app:normalize_function} compares $\sigma$ with alternative normalization functions and shows similar trends, indicating that the method is not sensitive to the particular bounded mapping.

\paragraph{Confidence Updating.}
Finally, we use the node-level trust weight $w_i^{(t)}$ to regulate how confidence evolves when an agent updates its response using peer references. Standard single-agent UQ treats confidence as a static property of the terminal output. In contrast, \methodname~tracks confidence along the interaction trajectory and allows it to change in proportion to the credibility of the peer evidence. We update dynamic confidence with a credibility-weighted moving average:
\begin{equation}
    c_{i,\mathrm{dyn}}^{(t)} = (1 - w_i^{(t)}) \cdot c_{i,\mathrm{dyn}}^{(t-1)} + w_i^{(t)} \cdot c_{i}^{(t)},
    \label{eq:calibrated_uncertainty}
\end{equation}
where $w_i^{(t)} \in [0,1]$ is the peer-trust weight, $c_i^{(t)}$ is the raw confidence of the current generation at node $v_i^{(t)}$, and $c_{i,\mathrm{dyn}}^{(t-1)}$ is the dynamic confidence from the previous round. This rule yields two intuitive regimes:
\begin{itemize}
    \item \textbf{Low peer credibility ($w_i^{(t)} \to 0$):} The current update is likely influenced by unreliable references. We therefore suppress the change and retain the historical estimate, $c_{i,\mathrm{dyn}}^{(t)} \approx c_{i,\mathrm{dyn}}^{(t-1)}$, preventing spurious overconfidence or distorted uncertainty.
    \item \textbf{High peer credibility ($w_i^{(t)} \to 1$):} The update is supported by credible references. We allow confidence to track the current response, $c_{i,\mathrm{dyn}}^{(t)} \approx c_{i}^{(t)}$, reflecting reliable correction from peer interactions.
\end{itemize}

\subsection{System-level Uncertainty Aggregation}
Multi-sampling, consistency-based UQ methods are prohibitively expensive for MAS because they would require repeated sampling at each interaction node. To remain cost-neutral with respect to a completed MAS run, \methodname~uses only the single-response confidence signals already produced during generation.

The objective of \methodname~is to produce one uncertainty score for the entire MAS generation trajectory. Given an interaction graph $\mathcal{G}=(\mathcal{V},\mathcal{E})$, input $x$, and generated contents $\mathcal{Y}=\{g(v)\mid v\in\mathcal{V}\}$, we define system-level uncertainty as the negative average dynamic confidence over the trajectory:
\begin{equation}
    U_{\mathcal{G},\mathrm{dyn}}(\mathcal{Y}\mid x) \;=\; -\frac{1}{|\mathcal{V}|}\sum_{v_i^{(t)}\in\mathcal{V}} c_{i,\mathrm{dyn}}^{(t)},
    \label{eq:umas}
\end{equation}
We use the average in Eq.~\ref{eq:umas} because it yields a stable, length-normalized summary across MAS runs with different numbers of agents and rounds, and prevents any single interaction node from dominating the system score.

\subsection{\methodfix: Disentangling Aggregation from Credibility Updating}
\methodname~consists of two cost-neutral components: (i) aggregating uncertainty over the entire MAS interaction graph rather than only the terminal output, and (ii) a heuristic update of confidence along the trajectory using peer credibility. To isolate these contributions, we define a simpler structure-aware instantiation, \methodfix. It uses the same trajectory aggregation as Eq.~\ref{eq:umas}, but replaces the credibility-updated confidence $c_{i,\mathrm{dyn}}^{(t)}$ with the raw node confidence $c_i^{(t)}$; equivalently, it skips the update in Eq.~\ref{eq:calibrated_uncertainty}. Thus, \methodfix~is structure-aware but credibility-agnostic. Comparing the resulting settings disentangles the two sources of improvement:
\begin{itemize}
    \item \methodfix~vs.\ a single-output UQ baseline measures the gain from \emph{trajectory-level aggregation alone}.
    \item \methodname~vs.\ \methodfix~measures the additional gain from \emph{credibility-aware updating}.
\end{itemize}
Our experiments report both deltas and find that each contributes substantially. Because all quantities are computed from logits and graph structure already produced in a standard MAS run, \methodname~requires no extra generations, model calls, or auxiliary models. The complete pseudocode of \methodname~is provided in Algorithm~\ref{alg:umas} (Appendix~\ref{app:algorithm}).

\begin{table}[!t]
\centering
\caption{AUROC (\%) results. ``Ave'' represents the mean over the ten dataset cells in each row (GSM-H = GSM-Hard). Best results per column are shown in \textbf{bold}; second-best are \underline{underlined}.}
\label{tab:ue-auroc}
\begin{sc}
\footnotesize
\setlength{\tabcolsep}{3pt}
\resizebox{\linewidth}{!}{%
\begin{tabular}{@{}lccccccccccc@{}}
\toprule
\multicolumn{12}{c}{\textbf{Llama 4 Scout}} \\
\midrule
& \multicolumn{5}{c}{\textbf{Debate}} & \multicolumn{5}{c}{\textbf{DyLAN}} & \\
\cmidrule(lr){2-6} \cmidrule(lr){7-11}
\textbf{Method} & \textbf{MMLUP} & \textbf{AQUA} & \textbf{MedMC} & \textbf{MBPP} & \textbf{GSM-H} &
\textbf{MMLUP} & \textbf{AQUA} & \textbf{MedMC} & \textbf{MBPP} & \textbf{GSM-H} & \textbf{Ave} \\
\midrule
PPL            & 67.67 & 64.48 & 56.53 & 60.03 & 36.02 & 60.15 & 71.94 & 53.93 & \textbf{58.61} & 55.28 & 58.46 \\
MSP            & 70.70 & 78.61 & 60.43 & 62.02 & 47.98 & 61.02 & 75.00 & 54.95 & \underline{58.39} & 57.47 & 62.66 \\
PE             & 69.58 & 70.33 & 57.81 & 59.99 & 37.09 & 63.58 & 74.03 & 54.77 & 52.53 & 56.44 & 59.62 \\
SE             & 47.81 & 47.33 & 45.83 & 37.42 & 51.41 & 50.29 & 55.30 & 56.09 & 48.09 & 49.41 & 48.90 \\
SentSAR        & 57.86 & 74.52 & 58.18 & 59.59 & 50.59 & 54.35 & 49.03 & 48.09 & 51.74 & 48.82 & 55.28 \\
Deg            & 54.89 & 67.75 & 59.20 & \underline{64.29} & 47.77 & 54.22 & 65.48 & 52.03 & 54.40 & \underline{78.01} & 59.80 \\
CoCoA          & 68.91 & \underline{79.39} & 60.74 & 62.36 & 56.13 & 60.94 & 73.02 & 54.70 & 58.27 & 55.36 & 62.98 \\
G-NLL          & 66.95 & 77.33 & 61.23 & 60.47 & 46.97 & 62.31 & 70.81 & 55.24 & 55.81 & 60.18 & 61.73 \\
\midrule
SE$^*$         & 55.08 & 50.70 & 48.30 & 52.06 & 55.80 & 56.99 & 67.02 & 60.50 & 55.06 & 53.47 & 55.50 \\
SentSAR$^*$    & 67.73 & 67.49 & 55.49 & 59.29 & 42.70 & 56.54 & 68.15 & 50.21 & 49.76 & 57.10 & 57.45 \\
Deg$^*$        & 58.82 & 62.76 & 58.02 & 62.63 & 62.56 & 55.62 & 72.24 & 52.68 & 37.36 & 60.62 & 58.33 \\
CoCoA$^*$      & 69.20 & 78.59 & 60.43 & 62.43 & 49.53 & 61.58 & 76.62 & 54.37 & 58.04 & 59.24 & 63.00 \\
\midrule
\methodfix     & \textbf{80.97} & 79.17 & \underline{72.68} & 62.00 & \underline{64.25} & \underline{79.71} & \underline{80.11} & \underline{66.41} & 51.93 & 65.97 & \underline{70.32} \\
\methodname    & \underline{80.90} & \textbf{79.77} & \textbf{74.38} & \textbf{64.56} & \textbf{72.69} & \textbf{83.70} & \textbf{80.63} & \textbf{76.16} & 56.98 & \textbf{79.48} & \textbf{74.93} \\
\midrule\midrule
\multicolumn{12}{c}{\textbf{GPT-4o mini}} \\
\midrule
& \multicolumn{5}{c}{\textbf{Debate}} & \multicolumn{5}{c}{\textbf{DyLAN}} & \\
\cmidrule(lr){2-6} \cmidrule(lr){7-11}
\textbf{Method} & \textbf{MMLUP} & \textbf{AQUA} & \textbf{MedMC} & \textbf{MBPP} & \textbf{GSM-H} &
\textbf{MMLUP} & \textbf{AQUA} & \textbf{MedMC} & \textbf{MBPP} & \textbf{GSM-H} & \textbf{Ave} \\
\midrule
PPL            & 60.12 & 51.43 & 58.67 & 47.85 & 47.83 & 61.50 & 53.88 & 66.07 & 41.93 & 45.30 & 53.46 \\
MSP            & 70.50 & 63.38 & 62.64 & 52.16 & 57.93 & 68.86 & 70.98 & 68.58 & 47.49 & 52.64 & 61.52 \\
PE             & 74.06 & 69.83 & 60.51 & 48.16 & 55.97 & 72.30 & 75.51 & 70.37 & 46.69 & 63.28 & 63.67 \\
SE             & 54.13 & 53.49 & 57.73 & 46.93 & 55.14 & 46.87 & 53.45 & 43.73 & 42.73 & 44.05 & 49.83 \\
SentSAR        & 75.42 & 71.45 & 62.79 & 51.79 & 60.16 & 65.59 & 69.38 & 71.08 & 46.63 & 51.79 & 62.61 \\
Deg            & 56.76 & 61.76 & 45.68 & 52.90 & 55.90 & 72.95 & 79.45 & 58.63 & 53.35 & \underline{76.28} & 61.37 \\
CoCoA          & 67.48 & 66.36 & 58.93 & 51.43 & 60.73 & 68.82 & 67.54 & 66.19 & 46.65 & 49.28 & 60.34 \\
G-NLL          & 70.62 & 60.86 & 62.04 & 53.04 & 59.19 & 68.72 & 70.98 & 68.58 & 47.49 & 52.64 & 61.42 \\
\midrule
SE$^*$         & 57.86 & 61.86 & 56.06 & 53.27 & 56.18 & 55.34 & 59.81 & 46.12 & 45.94 & 49.76 & 54.22 \\
SentSAR$^*$    & \textbf{77.60} & 71.23 & 62.30 & 53.07 & 59.72 & 73.32 & 73.87 & \underline{72.73} & 47.80 & 57.00 & 64.86 \\
Deg$^*$        & 56.13 & 66.15 & 46.58 & 55.40 & 60.94 & 58.75 & 67.05 & 56.49 & 47.80 & 69.35 & 58.46 \\
CoCoA$^*$      & 71.39 & 65.47 & 57.28 & 52.18 & 58.13 & 74.83 & 70.08 & 66.49 & 47.14 & 57.45 & 62.04 \\
\midrule
\methodfix     & \underline{77.42} & \underline{81.70} & \underline{66.03} & \underline{57.62} & \underline{67.89} & \underline{75.58} & \underline{84.22} & \textbf{73.17} & \underline{59.36} & 71.93 & \underline{71.49} \\
\methodname    & 74.40 & \textbf{85.51} & \textbf{71.03} & \textbf{60.37} & \textbf{74.50} & \textbf{76.54} & \textbf{87.48} & 72.33 & \textbf{61.69} & \textbf{76.39} & \textbf{74.02} \\
\bottomrule
\end{tabular}%
}
\end{sc}
\vspace{-0.5em}
\end{table}

\section{Experiments}\label{sec:experiments}

\subsection{Experimental Setup}

\paragraph{Benchmarks.}
We evaluate on five benchmarks spanning distinct cognitive domains: MMLU-Pro (general knowledge)~\cite{wang2024mmlu}, AQUA-RAT (mathematics)~\cite{ling2017program}, MedMCQA (medicine)~\cite{pal2022medmcqa}, MBPP (programming)~\cite{austin2021program}, and GSM-Hard (mathematics)~\cite{gao2022pal}. Specifications are in Appendix~\ref{app:dataset}.

\paragraph{Multi-Agent Systems and Backbone Models.}
To evaluate \methodname~on collaborative update mechanisms, we integrate it into two representative architectures: Debate~\cite{du2023improving}, a foundational paradigm for iterative peer critique, and DyLAN~\cite{liu2024dynamic}, an adaptive framework that dynamically optimizes team composition. We employ two distinct state-of-the-art LLMs as backbone agents: GPT-4o mini~\cite{achiam2023gpt}, a cost-efficient commercial model, and Llama 4 Scout~\cite{meta2025llama4}, a leading open-source Mixture-of-Experts model. Detailed architectural specifications are provided in Appendix~\ref{app:mas}.

\paragraph{UQ Baselines.}
We benchmark \methodname~against eight prominent Uncertainty Quantification methodologies originally developed for single-turn QA tasks: Perplexity (PPL)~\cite{fomicheva2020unsupervised}, MSP, PE~\cite{gawlikowski2023survey}, SE~\cite{kuhn2023semantic}, SentSAR~\cite{duan2024shifting}, Deg~\cite{lin2023generating}, CoCoA~\cite{vashurin2025uncertaintyquantificationllmsminimum}, and G-NLL~\cite{aichberger2024rethinking}. For all baselines, uncertainty is computed based on the final output of the MAS; we discuss this choice and provide an extended comparison where consistency-based baselines additionally aggregate across intermediate generations in Appendix~\ref{app:baselines_intermediate}.

A critical challenge in applying consistency-based methods to MAS is the prohibitive computational cost; these methods typically analyze semantic equivalence across multiple stochastic samples, which becomes resource-intensive given the extensive length of MAS trajectories. To address this bottleneck, we implement an optimization strategy that computes semantic consistency over a condensed representation of the generation, rather than the raw output (we denote methods using this optimization with an asterisk, e.g., SE$^*$; detailed implementation is provided in Appendix~\ref{app:star_method}). For all methods requiring multi-sampling, we standardize the sample number to $5$.

\paragraph{Evaluation Metric.}Following prior UQ work~\cite{kuhn2023semantic}, we evaluate the efficacy of \methodname~using the Area Under the Receiver Operating Characteristic Curve (AUROC), which measures the alignment between the estimated uncertainty and the correctness of the MAS's response. Correctness is determined by exact match between the extracted answer and the ground-truth answer for the multiple-choice and short-answer tasks (MMLU-Pro, AQUA-RAT, MedMCQA, GSM-Hard), and by pass@1 over all provided test cases for the code task (MBPP). Results are robust to the correctness definition and to response length (Appendices~\ref{app:extraction} and~\ref{app:length}). We additionally report Expected Calibration Error (ECE) in Appendix~\ref{app:ece} to verify that the improvements extend beyond ranking to calibration.

\subsection{Main Result}
Table~\ref{tab:ue-auroc} reports AUROC for \methodname~against established baselines on a random 200-instance subset per dataset. Both \methodfix~and \methodname~substantially improve uncertainty reliability across architectures, datasets, and backbones: on Llama 4 Scout the best baseline reaches Ave $63.00$, while \methodfix~and \methodname~reach $70.32$ and $74.93$; on GPT-4o mini the best baseline reaches $64.86$, versus $71.49$ and $74.02$, respectively. The two design choices decompose cleanly: trajectory-level aggregation alone (\methodfix~vs.\ best baseline) contributes $+7.32$/$+6.63$ AUROC across the two backbones, and credibility-aware updating (\methodname~vs.\ \methodfix) adds a further $+4.61$/$+2.53$; aggregation is thus the primary source of improvement. The credibility-aware contribution is most pronounced when peer confidences span a wide range (e.g., GSM-Hard, MedMCQA) and shrinks when peers are uniformly confident, consistent with the intuition that peer weighting matters most when peer reliability varies. Appendix~\ref{app:qwen} reproduces the trend on a third backbone (Qwen 3.5-9B~\cite{qwen3.5}); case studies are in Appendix~\ref{app:case}.

\subsection{Selective Prediction}
Beyond standard UQ metrics, we evaluate the practical utility of \methodname~in selecting the optimal response from multiple candidate generations~\cite{geifman2017selective}. Specifically, for each input $x$, we generate $K=5$ independent response trajectories $\{\gY_1, \dots, \gY_K\}$ using a temperature setting of $0.5$. The final system output is determined by selecting the trajectory that minimizes the estimated uncertainty:

\begin{equation*}
    \hat{y} = \gY_{k^*} \quad \text{where} \quad k^* = \argmin_{i \in \{1,\dots,K\}} U_{\gG}(\gY_i \mid x).
\end{equation*}

Table~\ref{tab:optimize_system} reports the overall accuracy of this selection mechanism using GPT-4o mini. The results indicate that our method significantly enhances system performance compared to the original MAS, with absolute accuracy gains under \methodname~ranging from 5.5 points (DyLAN-MMLUP) up to 12.6 points (Debate-AQUA). Notably, \methodname~or \methodfix~achieves the highest selection accuracy in every cell evaluated, and never falls below the strongest UQ baseline.

\begin{table}[!hp]
\centering
\caption{Comparative analysis of selective prediction accuracy (\%) across different UQ methods. The \textit{Original} denotes the standard performance of the MAS relying on a single generation trajectory, without applying any selection mechanism. Per row: \textbf{best} / \underline{second-best}.}
\label{tab:optimize_system}
\begin{sc}
\setlength{\tabcolsep}{2.5pt}
\resizebox{\linewidth}{!}{%
\begin{tabular}{@{}ll c cccc cc cc cc cc cc@{}}
\toprule
\textbf{Dataset} & \textbf{MAS} & \textit{Orig.} & PPL & MSP & G-NLL & PE & SE & SE$^*$ & SentSAR & SentSAR$^*$ & Deg & Deg$^*$ & CoCoA & CoCoA$^*$ & \methodfix & \methodname \\
\midrule
\multirow{2}{*}{MMLUP}
 & Debate & 70.0 & 74.0 & 73.5 & 74.5 & 75.0 & 72.0 & 72.0 & 73.0 & 74.5 & 68.5 & \underline{75.0} & 73.5 & 74.5 & 74.5          & \textbf{76.0} \\
 & DyLAN & 71.0 & 66.0 & 68.5 & 68.5 & 74.0 & 71.0 & 72.0 & 71.5 & 71.0 & \underline{76.5} & 74.5 & 72.5 & 75.0 & \textbf{77.0} & \underline{76.5} \\
\midrule
\multirow{2}{*}{AQUA}
 & Debate & 83.7 & 85.3 & 91.1 & 91.1 & 94.2 & 81.6 & 83.7 & 93.2 & 93.7 & 93.2 & 90.0 & 90.5 & 92.1 & \underline{94.8} & \textbf{96.3} \\
 & DyLAN & 83.2 & 81.6 & 89.5 & 89.5 & 90.0 & 83.2 & 83.7 & 90.0 & 91.1 & 89.5 & 81.1 & 87.4 & 83.2 & \underline{92.2} & \textbf{93.7} \\
\bottomrule
\end{tabular}%
}
\end{sc}
\end{table}

\subsection{Ablation Study}

\begin{figure}[!t]
    \centering
    \includegraphics[width=0.95\textwidth]{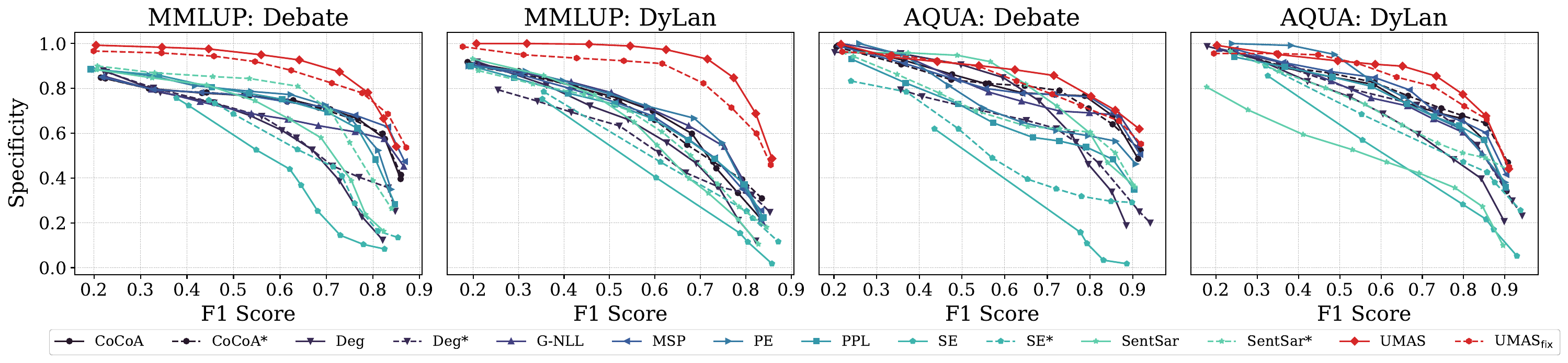}
    \caption{Trade-off analysis between F1 Score and Specificity on the test dataset using Llama 4 Scout. The curves illustrate performance variations across a range of classification thresholds $\delta \in \{0.1, 0.2, \dots, 0.9\}$. Methods exhibiting trajectories closer to the upper-right corner demonstrate superior efficacy in hallucination detection.}
    \label{fig:hallucination_detection}
    \vspace{-4mm}
\end{figure}

\paragraph{Hallucination Detection.}
A primary application of UQ is automated hallucination detection in LLM generations~\cite{maynez-etal-2020-faithfulness,manakul2023selfcheckgpt}. When ground truth is available, we define a hallucination as an incorrect response that does not match the reference answer. Accordingly, hallucination detection is a binary classification task: a response is flagged as hallucinated if its estimated uncertainty exceeds a threshold $\gamma$. To determine the optimal $\gamma$, a threshold-selection step is required. In our experiments, we utilize a 20\% validation set ($N=40$) from all data to derive a threshold for each method. The objective is to maximize specificity (True Negative Rate) while maintaining a target F1 score, formalized as:

\begin{equation*}
    \gamma^* = \operatorname*{argmax}_{\gamma} \text{Specificity}(\gamma)
    \quad \text{s.t.} \quad
    \text{F1}(\gamma) \ge \delta,
    \label{eq:threshold_optimization}
\end{equation*}

where $\delta$ is the target F1 level. Figure~\ref{fig:hallucination_detection} reports specificity across varying F1 constraints on the test set. \methodfix~and \methodname~consistently outperform all baselines, with \methodname~closer to the upper-right corner than \methodfix~in most settings, confirming the benefit of credibility-aware uncertainty dynamics.

\begin{figure}[!t]
    \centering
    \includegraphics[width=0.95\textwidth]{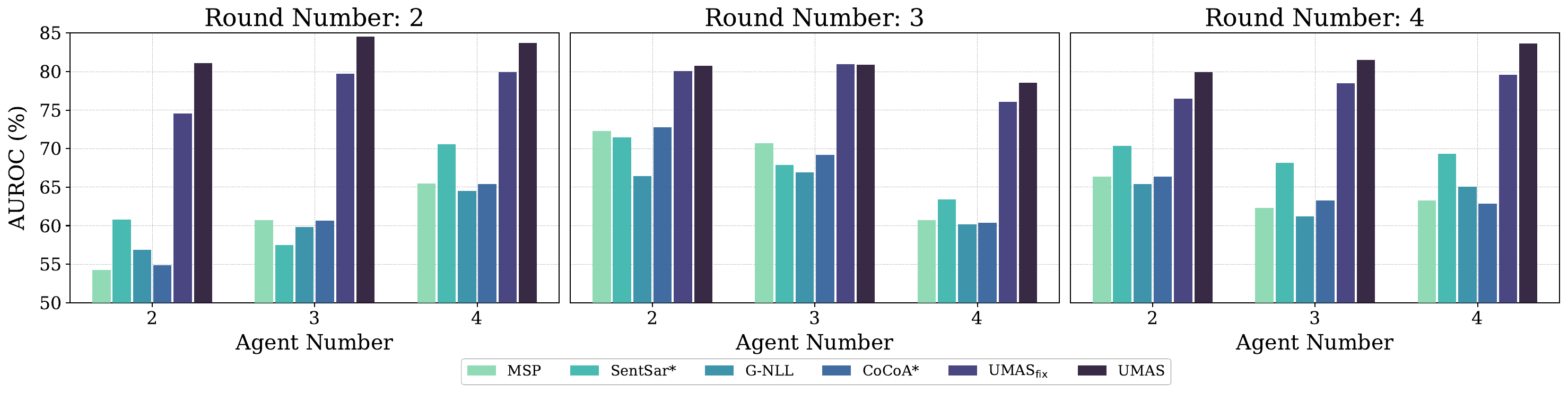}
    \caption{Comparative AUROC performance (\%) across varying agent system scales using Llama 4 Scout.}
    \label{fig:mas_scale}
    \vspace{-5mm}
\end{figure}

\paragraph{MAS Scalability.}
Multi-Agent Systems exhibit inherent architectural flexibility, allowing for variable configurations of agent count $N$ and communication rounds $M$. It is critical to assess UQ methods across these varying scales. Figure~\ref{fig:mas_scale} presents a comparative analysis between \methodname~and the four top-performing baselines across a comprehensive range of configurations, specifically $N, M \in \{2, 3, 4\}$. The results show that \methodfix~and \methodname~consistently outperform all baselines across the evaluated scales, demonstrating strong generalizability across diverse MAS deployments. Overall, \methodname~achieves the best performance in all settings.

\begin{figure}[!t]
    \centering
    \includegraphics[width=\linewidth]{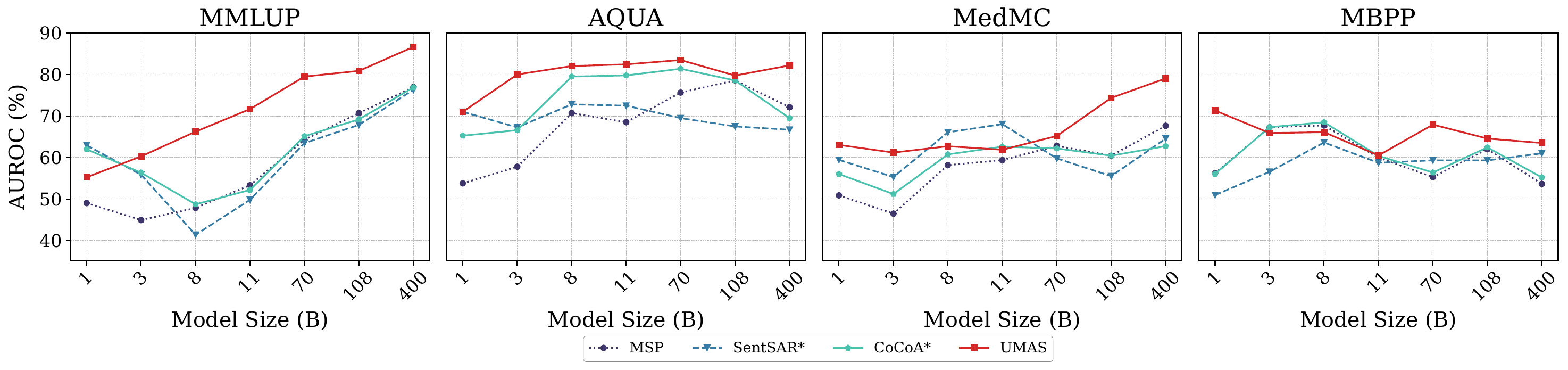}
    \caption{Comparative AUROC performance (\%) across varying agent parameter scales.}
    \label{fig:agent_scale}
    \vspace{-5mm}
\end{figure}

\paragraph{Agent Scalability.}
We evaluate \methodname~across seven Llama variants ranging from 1B to 400B total parameters: Llama-3.2 (1B, 3B, 11B-Vision)~\cite{dubey2024llama}, Llama-3.1-8B~\cite{dubey2024llama}, Llama-3.3-70B~\cite{dubey2024llama}, Llama 4 Scout (109B)~\cite{meta2025llama4}, and Llama 4 Maverick (400B)~\cite{meta2025llama4}. For models below 8B, we use MMLU~\cite{hendrycks2020measuring} since MMLU-Pro~\cite{wang2024mmlu} exceeds their reasoning capability. Figure~\ref{fig:agent_scale} reports AUROC across these seven scales on four datasets; \methodname~consistently outperforms most baselines, confirming scalability across model sizes.

\paragraph{Impact of Confidence Metrics.}\label{para:alternative_confidence}
\methodname~is metric-agnostic: any strictly positive single-pass uncertainty score $u$ can be plugged in as $c=-u$ (for MSP, $u=-\log P$). Replacing MSP with PE or PPL for the node-level confidence score (Llama 4 Scout, Debate), we observe substantial gains over the raw baseline metric in 11 of 12 configurations (Table~\ref{tab:ue_different_baselines}), with the sole exception of PPL on MBPP, where \methodname~is essentially on par with the baseline (59.87 vs.\ 60.03). The framework therefore inherits future single-turn UQ advances at no architectural cost.

\begin{table}[!htp]
\centering
\caption{Comparative AUROC performance (\%) utilizing distinct confidence metrics using Llama 4 Scout and Debate. ``Base'' is the raw confidence metric; \methodname~applies our framework on top.}
\label{tab:ue_different_baselines}
\begin{sc}
\setlength{\tabcolsep}{3pt}
\resizebox{\linewidth}{!}{%
\begin{tabular}{@{}l cccc cccc cccc@{}}
\toprule
& \multicolumn{4}{c}{\textbf{MSP}} & \multicolumn{4}{c}{\textbf{PPL}} & \multicolumn{4}{c}{\textbf{PE}} \\
\cmidrule(lr){2-5} \cmidrule(lr){6-9} \cmidrule(lr){10-13}
\textbf{Method} & MMLUP & AQUA & MedMC & MBPP & MMLUP & AQUA & MedMC & MBPP & MMLUP & AQUA & MedMC & MBPP \\
\midrule
Base       & 70.70          & 78.61          & 60.43          & 62.02          & 67.67          & 64.48          & 56.53          & \textbf{60.03} & 69.58          & 70.33          & 57.81          & 59.99          \\
\methodname & \textbf{80.90} & \textbf{79.77} & \textbf{74.38} & \textbf{64.56} & \textbf{77.98} & \textbf{72.16} & \textbf{70.88} & 59.87          & \textbf{79.01} & \textbf{73.69} & \textbf{70.29} & \textbf{60.67} \\
\bottomrule
\end{tabular}%
}
\end{sc}
\end{table}

\paragraph{Computational Efficiency.}\leavevmode\par\nobreak
\noindent\begin{minipage}[t]{0.40\linewidth}
\raggedright
Since MAS already incurs $N \times M$ inference calls per query, the UQ mechanism's overhead matters. \methodname~uses only single-trajectory logits and lightweight CPU-bound probability computations, avoiding the latency and hardware dependencies of multi-sample consistency methods (e.g., SE, SentSAR with DeBERTa-large entailment). Table~\ref{tab:efficiency} compares costs against baselines (hardware in Appendix~\ref{app:env}), where \textit{In.} denotes the inference latency and \textit{Un.} represents the time cost associated with uncertainty estimation.
\end{minipage}\hfill\begin{minipage}[t]{0.56\linewidth}
\vspace{-1.5em}
\centering
{\footnotesize\captionof{table}{Comparison of computational efficiency. Latency and AUROC measured on Llama 4 Scout under Debate on MMLU-Pro.}\label{tab:efficiency}}
\vspace{-0.5em}
\scriptsize
\setlength{\tabcolsep}{2.5pt}
\begin{sc}
\resizebox{\linewidth}{!}{%
\begin{tabular}{lcccc}
\toprule
\textbf{Method} & \textbf{In. (s)} & \textbf{Un. (ms)} & \textbf{GPU Free} & \textbf{AUROC}\\
\midrule
\multicolumn{5}{l}{\textit{Multi-Sampling}} \\
SE & 127.32 & 1813.92 & \ding{55} & 47.81\\
SentSAR & 127.32 & 263.30 & \ding{55} & 57.86\\
Deg & 127.32 & 3726.16 & \ding{55} & 54.89\\
CoCoA & 127.32 & 262.85 & \ding{55} & 68.91\\
G-NLL & 127.32 & 0.181 & \checkmark & 66.95\\
PE & 127.32 & 0.164 & \checkmark & 69.58\\
\midrule
\multicolumn{5}{l}{\textit{Single Sampling}} \\
PPL & 84.17 & 0.046 & \checkmark & 67.67\\
MSP & 84.17 & 0.021 & \checkmark & 70.70\\
\methodname & 84.17 & 0.143 & \checkmark & \textbf{80.90}\\
\bottomrule
\end{tabular}
}
\end{sc}

\end{minipage}

\section{Conclusion}

We show that standard single-agent UQ can become unreliable when applied to the final output of MAS, since it ignores trajectory-level interactions and influence credibility. We propose \methodname, a cost-neutral, trajectory-level UQ framework that aggregates node uncertainties over the interaction graph. \methodfix~yields strong gains via simple aggregation, and \methodname~further improves reliability by estimating the credibility of inter-agent influence from each generation's intrinsic confidence. Two design choices drive the gains: aggregating uncertainty over the full interaction graph rather than the terminal output (the larger contribution), and weighting peer evidence by its intrinsic confidence (the additional gain, most pronounced when peer reliability varies). Across MAS paradigms, backbones, and benchmarks, \methodname~consistently improves AUROC and benefits selective prediction and hallucination detection without extra sampling.

\clearpage

\bibliographystyle{plainnat}
\bibliography{reference}

\newpage
\appendix
\onecolumn
\raggedbottom

\section{Limitations}\label{app:limitations}

We acknowledge three limitations of \methodname.
(i) \methodname~requires access to token-level logits, making it hard to apply on fully black-box APIs that expose only completion text.
(ii) Our evaluation focuses on collaborative MAS protocols (Debate, DyLAN); adversarial MAS architectures, where agents may strategically misreport confidence, remain unexplored. More fundamentally, the credibility rule assumes cooperative peers: in role-asymmetric protocols such as proposer--critic, a critic's confidence cannot be interpreted without the semantic content and role of its critique.
(iii) When all agents are systematically and confidently wrong --- i.e., they share a common high-confidence misconception --- peer credibility signals cannot detect the failure, since no peer disagrees with the consensus.

\section{Broader Impact}\label{app:broader_impact}

\methodname~targets a reliability gap in deployed multi-agent LLM systems: detecting erroneous responses without incurring additional inference cost. More accurate uncertainty estimates support trustworthy deployment in high-stakes domains (e.g., medical question answering, code generation), where flagging unreliable outputs is essential. Two considerations bear on responsible use. First, like any uncertainty estimator, \methodname~can equally support legitimate quality assurance and adversarial probing of system weaknesses; we do not anticipate it unlocking misuse beyond what is already achievable via standard single-agent UQ. Second, \methodname~estimates self-reported uncertainty, not factual correctness --- a low-uncertainty response should not be interpreted as a guarantee of accuracy, and uncertainty quantification complements but does not replace ground-truth verification.

\section{Multi-Agent Systems}\label{app:mas}
In this section, we provide a detailed specification of the Multi-Agent Systems (MAS) utilized in this study.

\subsection{Architecture}

\begin{figure}[htbp]
    \centering
    \begin{subfigure}[b]{0.45\textwidth}
        \centering
        \includegraphics[width=\textwidth]{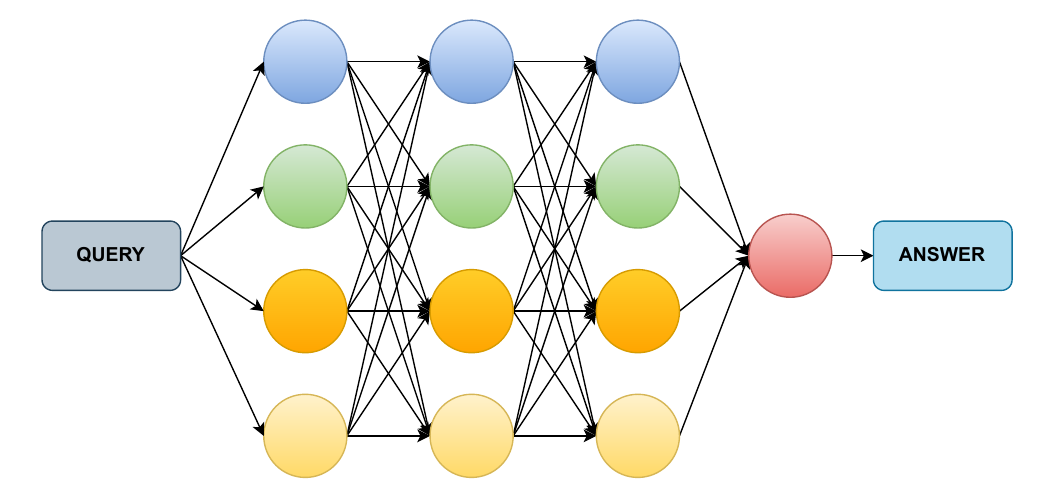}
        \caption{Illustration for Debate System}
        \label{fig:debate_illustration}
    \end{subfigure}
    \hspace{-0.2cm}
    \begin{subfigure}[b]{0.45\textwidth}
        \centering
        \includegraphics[width=\textwidth]{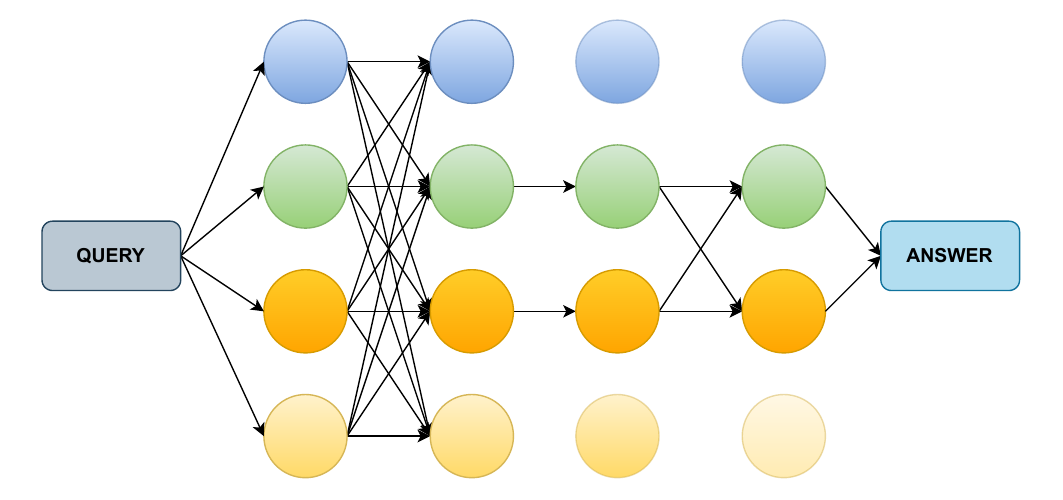}
        \caption{Illustration for DyLAN System}
        \label{fig:dylan_illustration}
    \end{subfigure}

    \caption{Architectural illustrations of the two MAS paradigms used in our experiments. \textbf{(a)} The Debate system, in which agents iteratively read, critique, and revise their responses based on peers. \textbf{(b)} The DyLAN system, which dynamically selects the most capable contributors via an intermediate ranker.}
    \label{fig:both_images}
\end{figure}

\paragraph{Debate.}
The Debate framework~\cite{du2023improving} is inspired by the "Society of Mind" theory. The process initiates with agents independently generating candidate answers, followed by iterative rounds of reading, critiquing, and updating responses based on peer feedback. This recursive exchange enables agents to identify and rectify hallucinations or logical fallacies in their reasoning. Ultimately, the group converges on a unified, higher-accuracy consensus solely through prompting strategies and context sharing. The architecture is illustrated in Figure~\ref{fig:debate_illustration}.

\paragraph{DyLAN.}
The Dynamic LLM-Agent Network (DyLAN) framework is designed to transcend static, fixed-agent teams by dynamically selecting and organizing agents tailored to specific tasks. During the Task Solving stage, agents collaborate within a feed-forward structure that treats communication steps as layers in a network. This configuration facilitates an evolving communication topology, utilizing an intermediate ranker to continuously assess contribution quality and deactivate less effective agents in real-time. This mechanism ensures that the reasoning process is driven exclusively by the most capable contributors. The overall architecture is depicted in Figure~\ref{fig:dylan_illustration}. \\
Unless otherwise specified, all experiments in this paper use a Debate system with 3 agents and 3 rounds. For the DyLAN system, we use 4 agents and 3 rounds.

\subsection{Task Accuracy}\label{app:mas_accuracy}

To contextualize the AUROC results in the main paper, we report the underlying task accuracy of each (backbone, MAS) pair on every benchmark in Table~\ref{tab:mas_accuracy}. \textbf{Base} denotes a single-agent inference baseline; for fair comparison, it is taken from the first agent's first-round response in the corresponding Debate run (which is generated independently before any peer communication, and thus uses identical decoding configuration to the MAS run). Across nearly every (backbone, dataset) pair, both Debate and DyLAN improve over Base, motivating MAS as a deployment paradigm and the need for MAS-specific UQ in the first place.

\begin{table}[h]
\centering
\caption{Task accuracy (\%) for each (backbone, MAS) configuration across the five evaluation benchmarks. ``Base'' is the first-round response of agent~1 in Debate (i.e., a single-agent baseline using identical decoding configuration); ``Debate'' uses 3 agents $\times$ 3 rounds; ``DyLAN'' uses 4 agents $\times$ 3 rounds. Per (backbone, dataset) column: \textbf{best} / \underline{second-best}.}
\label{tab:mas_accuracy}
\begin{sc}
\setlength{\tabcolsep}{6pt}
\begin{tabular}{@{}llccccc@{}}
\toprule
\textbf{Backbone} & \textbf{System} & \textbf{MMLU-Pro} & \textbf{AQUA} & \textbf{MedMCQA} & \textbf{MBPP} & \textbf{GSM-Hard} \\
\midrule
\multirow{3}{*}{Llama 4 Scout}
 & Base   & 70.00 & 85.79 & 71.00 & 54.50 & \underline{60.50} \\
 & Debate & \underline{75.50} & \underline{87.37} & \textbf{73.00} & \textbf{59.50} & \textbf{61.00} \\
 & DyLAN  & \textbf{77.00} & \textbf{88.42} & \underline{72.00} & \textbf{59.50} & 59.50 \\
\midrule
\multirow{3}{*}{GPT-4o mini}
 & Base   & 66.50 & 82.63 & 63.50 & 62.00 & 57.00 \\
 & Debate & \underline{70.00} & \textbf{83.68} & \underline{67.50} & \textbf{63.50} & \underline{59.50} \\
 & DyLAN  & \textbf{71.00} & \underline{83.16} & \textbf{68.00} & \underline{62.50} & \textbf{60.50} \\
\midrule
\multirow{3}{*}{Qwen 3.5-9B}
 & Base   & 82.50 & 88.95 & \textbf{69.50} & 60.50 & 67.68 \\
 & Debate & \underline{85.00} & \textbf{91.05} & \textbf{69.50} & \textbf{65.50} & \textbf{69.70} \\
 & DyLAN  & \textbf{86.41} & \underline{89.47} & 66.00 & \underline{64.50} & \underline{68.00} \\
\bottomrule
\end{tabular}
\end{sc}
\end{table}

\subsection{Prompt}

\begin{tcolorbox}[
    title={\large \textbf{Debate}},
    colback=gray!10,
    colframe=black,
    coltitle=white,
    fontupper=\rmfamily,
    boxrule=1.5pt,
    arc=2mm,
]

    \textbf{Initial Prompt:}

    \textcolor{query}{\{Query\}} Make sure to state your answer at the end of the response.
    \\

    \textbf{Update Prompt:}

    These are the recent/updated opinions from other agents:

    One agent response: \textcolor{reference}{\{Other Agent Response\}}

    One agent response: \textcolor{reference}{\{Other Agent Response\}}

    Use these opinions carefully as additional advice. Can you provide an updated answer? Make sure to state your answer at the end of the response.
    The original problem is \textcolor{query}{\{Query\}}. \\

    \textbf{Aggregation Prompt:}

    Task:
    \textcolor{query}{\{Query\}}

    Solution 1:
    \textcolor{reference}{\{Final Answer from Agent 1\}}

    Solution 2:
    \textcolor{reference}{\{Final Answer from Agent 2\}}

    Solution 3:
    \textcolor{reference}{\{Final Answer from Agent 3\}}

    Given all the above solutions, reason over them carefully and provide a final answer to the task.

\end{tcolorbox}

\begin{tcolorbox}[
    title={\large \textbf{DyLAN}},
    colback=gray!10,
    colframe=black,
    coltitle=white,
    fontupper=\rmfamily,
    boxrule=1.5pt,
    arc=2mm,
]

\textbf{Initial Prompt:}

You are a super-intelligent AI assistant capable of performing tasks more effectively than humans.\\
\textcolor{query}{\{Query\}}
\\

\textbf{Update Prompt:}

\textcolor{query}{\{Query\}}
\\

These are the responses from other agents:

Agent response 1: \textcolor{reference}{\{Other Agent Response\}}

Agent response 2: \textcolor{reference}{\{Other Agent Response\}}

Agent response 3: \textcolor{reference}{\{Other Agent Response\}}

Using the answer from other agents as additional advice with critical thinking, can you give an updated answer? Examine your solution and that other agents step by step. Notice that their answers might be all wrong. Please answer the question in detail. Along with the answer, give a score ranged from 1 to 5 to the solutions of other agents. Put all \textbf{N} scores in the form like [[1, 5, 2, ...]]. \\

\textbf{Ranking Prompt:}

\textcolor{query}{\{Query\}}
\\

These are the responses from other agents:

Agent response 1: \textcolor{reference}{\{Agent Response\}}

Agent response 2: \textcolor{reference}{\{Agent Response\}}

Agent response 3: \textcolor{reference}{\{Agent Response\}}

Agent response 4: \textcolor{reference}{\{Agent Response\}}

Please choose the best 2 answers and think step by step. Put your answer in the form like [1,2] or [3,4] at the end of your response.
\end{tcolorbox}

\subsection{Example}

\begin{xltabular}{\linewidth}{X}

    \caption{Representative MAS interaction trajectory under Debate, illustrating the verbosity of multi-agent generations across rounds.} \label{tab:long_example} \\
    \toprule
    \endfirsthead

    \caption[]{Representative MAS interaction trajectory (continued).} \\
    \endhead

    \multicolumn{1}{r}{\textit{Continued on next page...}} \\
    \endfoot

    \bottomrule
    \endlastfoot

    {\Large \textbf{Round 1}}  \newline
    {\textbf{Initial Prompt}} \newline
    \textit{Role}: User \newline
    \textit{Content}: {\small The following is a multiple-choice question:\newline Norcross agrees that if a being is incapable of moral reasoning, at even the most basic level, then it cannot be\newline \newline Choose the correct answer from the following options:\newline (A) an object of moral sympathy.\newline (B) a morally autonomous being.\newline (C) a moral patient.\newline (D) a being of moral integrity.\newline (E) a moral agent.\newline (F) a subject of moral discourse.\newline (G) an ethically responsible entity.\newline (H) a being of value.\newline (I) a moral observer.\newline (J) a morally accountable being. Make sure to state your answer at the end of the response.}
    \\
    \hline
    \textbf{Agent One Response} \newline
    {\small To solve this question, let's analyze the key terms and concepts related to moral philosophy:

    1. **Moral Agent**: A being capable of making moral judgments and decisions. This typically requires a level of moral reasoning.

    2. **Moral Patient**: A being that can be harmed or helped by the actions of others and is therefore a recipient of moral consideration. This does not necessarily require the ability to make moral judgments.

    3. **Moral Autonomy**: Refers to the ability of a being to make decisions based on its own moral principles. This implies a capacity for moral reasoning.

    4. **Moral Integrity**: This refers to the quality of being honest and having strong moral principles. It implies a certain level of moral maturity and reasoning.

    5. **Moral Sympathy/Object of Moral Sympathy**: This refers to the capacity to evoke feelings of sympathy or concern in others. It doesn't necessarily require the being to have moral reasoning capabilities.

    6. **Subject of Moral Discourse**: This refers to entities that can be discussed in terms of moral judgments. It can include both moral agents and patients.

    7. **Ethically Responsible Entity**: This implies a being capable of taking responsibility for its actions, which requires some level of moral reasoning.

    8. **Being of Value**: This simply means that the being has value and deserves consideration. It doesn't necessarily imply the capacity for moral reasoning.

    9. **Moral Observer**: A being that can observe and perhaps understand moral situations, but it doesn't necessarily imply the ability to make moral judgments.

    10. **Morally Accountable Being**: This refers to a being that can be held accountable for its actions, implying some level of moral reasoning.

    Given Norcross's statement that if a being is incapable of moral reasoning, at even the most basic level, then it cannot be a certain type of entity, we need to identify which of these terms necessarily implies the capacity for moral reasoning.

    - **(A) an object of moral sympathy**: Doesn't require moral reasoning.
    - **(B) a morally autonomous being**: Requires moral reasoning.
    - **(C) a moral patient**: Doesn't require moral reasoning.
    - **(D) a being of moral integrity**: Implies a level of moral maturity and thus reasoning.
    - **(E) a moral agent**: Directly implies the capacity for moral reasoning.
    - **(F) a subject of moral discourse**: Doesn't necessarily require moral reasoning.
    - **(G) an ethically responsible entity**: Implies some level of moral reasoning.
    - **(H) a being of value**: Doesn't require moral reasoning.
    - **(I) a moral observer**: While it might imply understanding, it doesn't directly imply making moral judgments.
    - **(J) a morally accountable being**: Implies the capacity for moral reasoning.

    The most direct and universally agreed-upon term that requires the capacity for moral reasoning is **(E) a moral agent**. Being a moral agent is fundamentally about having the capacity to make moral decisions, which directly involves moral reasoning.

    The best answer is E.} \\
    \hline
    \textbf{Agent Two Response} \newline
    {\small To solve this, let's analyze the terms given in the context of moral philosophy:

    - **Moral Agent**: An entity capable of making moral decisions and taking moral actions. This typically requires the ability to reason morally.
    - **Moral Patient**: An entity that can be affected by moral actions and is deserving of moral consideration, but may not necessarily be capable of moral reasoning itself.
    - **Moral Autonomy**: The ability of an entity to make decisions based on its own moral principles, which implies a level of moral reasoning.
    - **Moral Integrity**: Refers to the quality of being honest and having strong moral principles. This concept is more about the character of a moral agent rather than a prerequisite for being considered within moral discourse.
    - **Moral Observer**: An entity that can observe and perhaps understand moral actions but may not necessarily make moral judgments.
    - **Moral Sympathy**: The capacity to feel pity or compassion for another's suffering, which doesn't necessarily require moral reasoning but does require some level of emotional or cognitive capability.
    - **Subject of Moral Discourse**: This refers to entities that can be talked about in terms of moral judgments, which includes both moral agents and patients.
    - **Ethically Responsible Entity**: An entity that can be held responsible for its actions, implying a level of moral agency.
    - **Morally Accountable Being**: Similar to an ethically responsible entity, this implies the ability to be held accountable for one's actions, suggesting moral agency.
    - **Being of Value**: This simply means an entity that has value, which can be intrinsic or instrumental, and does not necessarily imply the capacity for moral reasoning.

    Given that Norcross agrees if a being is incapable of moral reasoning, at even the most basic level, then it cannot be a certain type of entity, we need to identify which term necessarily requires moral reasoning.

    - **Moral Agent (E)**: Necessarily requires moral reasoning.
    - **Morally Autonomous Being (B)**: Requires moral reasoning.
    - **Ethically Responsible Entity (G)**: Implies moral reasoning.
    - **Morally Accountable Being (J)**: Implies moral reasoning.

    However, the question seems to seek a term that is most directly impacted by the inability to perform moral reasoning.

    The term **Moral Agent (E)** directly refers to an entity that can perform moral actions and make moral decisions, which inherently requires the capacity for moral reasoning.

    The best answer is E.} \\
    \hline
    \textbf{Agent Three Response} \newline
    {\small To solve this question, let's analyze the given statement and the options provided.

    The statement by Norcross implies that if a being cannot engage in moral reasoning, even at a basic level, it cannot be considered something specific in terms of moral philosophy. The key term here is "moral reasoning," which is closely related to the capacity for making moral judgments, being morally responsible, or acting as a moral entity.

    Let's define some key terms to narrow down the options:
    - **Moral Agent**: An entity capable of making moral decisions and acting based on those decisions. This requires a level of moral reasoning.
    - **Moral Patient**: An entity that can be harmed or helped, and whose well-being is considered in moral deliberations. This does not necessarily require the capacity for moral reasoning.
    - **Moral Autonomy**: The capacity to make decisions based on one's own free will, particularly in a moral context. This implies a level of moral reasoning.

    Given these definitions:
    - A being incapable of moral reasoning cannot be a **moral agent (E)**, as being a moral agent requires the capacity for moral reasoning to make and act on moral decisions.
    - It does not necessarily follow that it cannot be a **moral patient (C)**, as patients can be entities that are affected by actions but do not have to have the capacity for moral reasoning themselves.
    - Options like **morally autonomous being (B)**, **morally accountable being (J)**, and **ethically responsible entity (G)** also imply a level of moral reasoning or capacity for moral action.

    Considering these points, the most direct and accurate completion of Norcross's statement would relate to the capacity for moral action or decision-making.

    The best answer is E.}
    \\
    \midrule
    {\Large \textbf{Round 2}}  \\
    \hline
    \textbf{Agent One Prompt} \newline
    \textit{Role}: User \newline
    \textit{Content}: {\small The following is a multiple-choice question:\newline Norcross agrees that if a being is incapable of moral reasoning, at even the most basic level, then it cannot be\newline \newline Choose the correct answer from the following options:\newline (A) an object of moral sympathy.\newline (B) a morally autonomous being.\newline (C) a moral patient.\newline (D) a being of moral integrity.\newline (E) a moral agent.\newline (F) a subject of moral discourse.\newline (G) an ethically responsible entity.\newline (H) a being of value.\newline (I) a moral observer.\newline (J) a morally accountable being. Make sure to state your answer at the end of the response.} \\
    \textit{Role}: Assistant \newline
    \textit{Content}: {\small To solve this question, let's analyze the key terms and concepts related to moral philosophy:  1. **Moral Agent**: A being capable of making moral judgments and decisions. This typically requires a level of moral reasoning.  2. **Moral Patient**: A being that can be harmed or helped by the actions of others and is therefore a recipient of moral consideration. This does not necessarily require the ability to make moral judgments.  3. **Moral Autonomy**: Refers to the ability of a being to make decisions based on its own moral principles. This implies a capacity for moral reasoning.  4. **Moral Integrity**: This refers to the quality of being honest and having strong moral principles. It implies a certain level of moral maturity and reasoning.  5. **Moral Sympathy/Object of Moral Sympathy**: This refers to the capacity to evoke feelings of sympathy or concern in others. It doesn't necessarily require the being to have moral reasoning capabilities.  6. **Subject of Moral Discourse**: This refers to entities that can be discussed in terms of moral judgments. It can include both moral agents and patients.  7. **Ethically Responsible Entity**: This implies a being capable of taking responsibility for its actions, which requires some level of moral reasoning.  8. **Being of Value**: This simply means that the being has value and deserves consideration. It doesn't necessarily imply the capacity for moral reasoning.  9. **Moral Observer**: A being that can observe and perhaps understand moral situations, but it doesn't necessarily imply the ability to make moral judgments.  10. **Morally Accountable Being**: This refers to a being that can be held accountable for its actions, implying some level of moral reasoning.  Given Norcross's statement that if a being is incapable of moral reasoning, at even the most basic level, then it cannot be a certain type of entity, we need to identify which of these terms necessarily implies the capacity for moral reasoning.  - **(A) an object of moral sympathy**: Doesn't require moral reasoning. - **(B) a morally autonomous being**: Requires moral reasoning. - **(C) a moral patient**: Doesn't require moral reasoning. - **(D) a being of moral integrity**: Implies a level of moral maturity and thus reasoning. - **(E) a moral agent**: Directly implies the capacity for moral reasoning. - **(F) a subject of moral discourse**: Doesn't necessarily require moral reasoning. - **(G) an ethically responsible entity**: Implies some level of moral reasoning. - **(H) a being of value**: Doesn't require moral reasoning. - **(I) a moral observer**: While it might imply understanding, it doesn't directly imply making moral judgments. - **(J) a morally accountable being**: Implies the capacity for moral reasoning.  The most direct and universally agreed-upon term that requires the capacity for moral reasoning is **(E) a moral agent**. Being a moral agent is fundamentally about having the capacity to make moral decisions, which directly involves moral reasoning.  The best answer is E.
    } \\
    \textit{Role}: User \newline
    \textit{Content}: {\small These are the recent/updated opinions from other agents:

     One agent response: ```To solve this, let's analyze the terms given in the context of moral philosophy:

    - **Moral Agent**: An entity capable of making moral decisions and taking moral actions. This typically requires the ability to reason morally.
    - **Moral Patient**: An entity that can be affected by moral actions and is deserving of moral consideration, but may not necessarily be capable of moral reasoning itself.
    - **Moral Autonomy**: The ability of an entity to make decisions based on its own moral principles, which implies a level of moral reasoning.
    - **Moral Integrity**: Refers to the quality of being honest and having strong moral principles. This concept is more about the character of a moral agent rather than a prerequisite for being considered within moral discourse.
    - **Moral Observer**: An entity that can observe and perhaps understand moral actions but may not necessarily make moral judgments.
    - **Moral Sympathy**: The capacity to feel pity or compassion for another's suffering, which doesn't necessarily require moral reasoning but does require some level of emotional or cognitive capability.
    - **Subject of Moral Discourse**: This refers to entities that can be talked about in terms of moral judgments, which includes both moral agents and patients.
    - **Ethically Responsible Entity**: An entity that can be held responsible for its actions, implying a level of moral agency.
    - **Morally Accountable Being**: Similar to an ethically responsible entity, this implies the ability to be held accountable for one's actions, suggesting moral agency.
    - **Being of Value**: This simply means an entity that has value, which can be intrinsic or instrumental, and does not necessarily imply the capacity for moral reasoning.

    Given that Norcross agrees if a being is incapable of moral reasoning, at even the most basic level, then it cannot be a certain type of entity, we need to identify which term necessarily requires moral reasoning.

    - **Moral Agent (E)**: Necessarily requires moral reasoning.
    - **Morally Autonomous Being (B)**: Requires moral reasoning.
    - **Ethically Responsible Entity (G)**: Implies moral reasoning.
    - **Morally Accountable Being (J)**: Implies moral reasoning.

    However, the question seems to seek a term that is most directly impacted by the inability to perform moral reasoning.

    The term **Moral Agent (E)** directly refers to an entity that can perform moral actions and make moral decisions, which inherently requires the capacity for moral reasoning.

    The best answer is E.```

     One agent response: ```To solve this question, let's analyze the given statement and the options provided.

    The statement by Norcross implies that if a being cannot engage in moral reasoning, even at a basic level, it cannot be considered something specific in terms of moral philosophy. The key term here is "moral reasoning," which is closely related to the capacity for making moral judgments, being morally responsible, or acting as a moral entity.

    Let's define some key terms to narrow down the options:
    - **Moral Agent**: An entity capable of making moral decisions and acting based on those decisions. This requires a level of moral reasoning.
    - **Moral Patient**: An entity that can be harmed or helped, and whose well-being is considered in moral deliberations. This does not necessarily require the capacity for moral reasoning.
    - **Moral Autonomy**: The capacity to make decisions based on one's own free will, particularly in a moral context. This implies a level of moral reasoning.

    Given these definitions:
    - A being incapable of moral reasoning cannot be a **moral agent (E)**, as being a moral agent requires the capacity for moral reasoning to make and act on moral decisions.
    - It does not necessarily follow that it cannot be a **moral patient (C)**, as patients can be entities that are affected by actions but do not have to have the capacity for moral reasoning themselves.
    - Options like **morally autonomous being (B)**, **morally accountable being (J)**, and **ethically responsible entity (G)** also imply a level of moral reasoning or capacity for moral action.

    Considering these points, the most direct and accurate completion of Norcross's statement would relate to the capacity for moral action or decision-making.

    The best answer is E.```

     Use these opinions carefully as additional advice, can you provide an updated answer? Make sure to state your answer at the end of the response.
     The original problem is The following is a multiple-choice question:
     Norcross agrees that if a being is incapable of moral reasoning, at even the most basic level, then it cannot be

    Choose the correct answer from the following options:
    (A) an object of moral sympathy.
    (B) a morally autonomous being.
    (C) a moral patient.
    (D) a being of moral integrity.
    (E) a moral agent.
    (F) a subject of moral discourse.
    (G) an ethically responsible entity.
    (H) a being of value.
    (I) a moral observer.
    (J) a morally accountable being.} \\
    \textbf{Agent One Response} \newline
    {\small Given the analysis and definitions provided, the key concept here is the inability of a being to engage in moral reasoning at even the most basic level. This directly impacts its capacity to perform actions that require moral judgment, decision-making, and responsibility.

    - **Moral Agent (E)**: Directly requires the capacity for moral reasoning to make and act on moral decisions.
    - **Morally Autonomous Being (B)**: Implies a level of moral reasoning for self-governance.
    - **Morally Accountable Being (J)**: Suggests the ability to be held accountable, implying moral reasoning.
    - **Ethically Responsible Entity (G)**: Implies a level of moral reasoning for responsibility.

    However, among these, **Moral Agent (E)** is the most straightforward term that necessitates moral reasoning. Being a moral agent is fundamentally about the capacity to engage in moral reasoning to make decisions and act morally.

    The best answer is E.} \\

    \hline

    \textbf{Agent Two Prompt} \newline

    \textit{Role}: User \newline
    \textit{Content}: {\small The following is a multiple-choice question:\newline  Norcross agrees that if a being is incapable of moral reasoning, at even the most basic level, then it cannot be\newline \newline Choose the correct answer from the following options:\newline (A) an object of moral sympathy.\newline (B) a morally autonomous being.\newline (C) a moral patient.\newline (D) a being of moral integrity.\newline (E) a moral agent.\newline (F) a subject of moral discourse.\newline (G) an ethically responsible entity.\newline (H) a being of value.\newline (I) a moral observer.\newline (J) a morally accountable being. Make sure to state your answer at the end of the response.} \newline

    \textit{Role}: Assistant \newline
    \textit{Content}: {\small To solve this, let's analyze the terms given in the context of moral philosophy:  - **Moral Agent**: An entity capable of making moral decisions and taking moral actions. This typically requires the ability to reason morally. - **Moral Patient**: An entity that can be affected by moral actions and is deserving of moral consideration, but may not necessarily be capable of moral reasoning itself. - **Moral Autonomy**: The ability of an entity to make decisions based on its own moral principles, which implies a level of moral reasoning. - **Moral Integrity**: Refers to the quality of being honest and having strong moral principles. This concept is more about the character of a moral agent rather than a prerequisite for being considered within moral discourse. - **Moral Observer**: An entity that can observe and perhaps understand moral actions but may not necessarily make moral judgments. - **Moral Sympathy**: The capacity to feel pity or compassion for another's suffering, which doesn't necessarily require moral reasoning but does require some level of emotional or cognitive capability. - **Subject of Moral Discourse**: This refers to entities that can be talked about in terms of moral judgments, which includes both moral agents and patients. - **Ethically Responsible Entity**: An entity that can be held responsible for its actions, implying a level of moral agency. - **Morally Accountable Being**: Similar to an ethically responsible entity, this implies the ability to be held accountable for one's actions, suggesting moral agency. - **Being of Value**: This simply means an entity that has value, which can be intrinsic or instrumental, and does not necessarily imply the capacity for moral reasoning.  Given that Norcross agrees if a being is incapable of moral reasoning, at even the most basic level, then it cannot be a certain type of entity, we need to identify which term necessarily requires moral reasoning.  - **Moral Agent (E)**: Necessarily requires moral reasoning. - **Morally Autonomous Being (B)**: Requires moral reasoning. - **Ethically Responsible Entity (G)**: Implies moral reasoning. - **Morally Accountable Being (J)**: Implies moral reasoning.  However, the question seems to seek a term that is most directly impacted by the inability to perform moral reasoning.   The term **Moral Agent (E)** directly refers to an entity that can perform moral actions and make moral decisions, which inherently requires the capacity for moral reasoning. The best answer is E.} \\

    \textit{Role}: User \newline
    \textit{Content}: {\small These are the recent/updated opinions from other agents:

     One agent response: ```To solve this question, let's analyze the key terms and concepts related to moral philosophy:

    1. **Moral Agent**: A being capable of making moral judgments and decisions. This typically requires a level of moral reasoning.

    2. **Moral Patient**: A being that can be harmed or helped by the actions of others and is therefore a recipient of moral consideration. This does not necessarily require the ability to make moral judgments.

    3. **Moral Autonomy**: Refers to the ability of a being to make decisions based on its own moral principles. This implies a capacity for moral reasoning.

    4. **Moral Integrity**: This refers to the quality of being honest and having strong moral principles. It implies a certain level of moral maturity and reasoning.

    5. **Moral Sympathy/Object of Moral Sympathy**: This refers to the capacity to evoke feelings of sympathy or concern in others. It doesn't necessarily require the being to have moral reasoning capabilities.

    6. **Subject of Moral Discourse**: This refers to entities that can be discussed in terms of moral judgments. It can include both moral agents and patients.

    7. **Ethically Responsible Entity**: This implies a being capable of taking responsibility for its actions, which requires some level of moral reasoning.

    8. **Being of Value**: This simply means that the being has value and deserves consideration. It doesn't necessarily imply the capacity for moral reasoning.

    9. **Moral Observer**: A being that can observe and perhaps understand moral situations, but it doesn't necessarily imply the ability to make moral judgments.

    10. **Morally Accountable Being**: This refers to a being that can be held accountable for its actions, implying some level of moral reasoning.

    Given Norcross's statement that if a being is incapable of moral reasoning, at even the most basic level, then it cannot be a certain type of entity, we need to identify which of these terms necessarily implies the capacity for moral reasoning.

    - **(A) an object of moral sympathy**: Doesn't require moral reasoning.
    - **(B) a morally autonomous being**: Requires moral reasoning.
    - **(C) a moral patient**: Doesn't require moral reasoning.
    - **(D) a being of moral integrity**: Implies a level of moral maturity and thus reasoning.
    - **(E) a moral agent**: Directly implies the capacity for moral reasoning.
    - **(F) a subject of moral discourse**: Doesn't necessarily require moral reasoning.
    - **(G) an ethically responsible entity**: Implies some level of moral reasoning.
    - **(H) a being of value**: Doesn't require moral reasoning.
    - **(I) a moral observer**: While it might imply understanding, it doesn't directly imply making moral judgments.
    - **(J) a morally accountable being**: Implies the capacity for moral reasoning.

    The most direct and universally agreed-upon term that requires the capacity for moral reasoning is **(E) a moral agent**. Being a moral agent is fundamentally about having the capacity to make moral decisions, which directly involves moral reasoning.

    The best answer is E.```} \\
    {\small One agent response: ```To solve this question, let's analyze the given statement and the options provided.

    The statement by Norcross implies that if a being cannot engage in moral reasoning, even at a basic level, it cannot be considered something specific in terms of moral philosophy. The key term here is "moral reasoning," which is closely related to the capacity for making moral judgments, being morally responsible, or acting as a moral entity.

    Let's define some key terms to narrow down the options:
    - **Moral Agent**: An entity capable of making moral decisions and acting based on those decisions. This requires a level of moral reasoning.
    - **Moral Patient**: An entity that can be harmed or helped, and whose well-being is considered in moral deliberations. This does not necessarily require the capacity for moral reasoning.
    - **Moral Autonomy**: The capacity to make decisions based on one's own free will, particularly in a moral context. This implies a level of moral reasoning.

    Given these definitions:
    - A being incapable of moral reasoning cannot be a **moral agent (E)**, as being a moral agent requires the capacity for moral reasoning to make and act on moral decisions.
    - It does not necessarily follow that it cannot be a **moral patient (C)**, as patients can be entities that are affected by actions but do not have to have the capacity for moral reasoning themselves.
    - Options like **morally autonomous being (B)**, **morally accountable being (J)**, and **ethically responsible entity (G)** also imply a level of moral reasoning or capacity for moral action.

    Considering these points, the most direct and accurate completion of Norcross's statement would relate to the capacity for moral action or decision-making.

    The best answer is E.```} \\
    {\small Use these opinions carefully as additional advice, can you provide an updated answer? Make sure to state your answer at the end of the response.
     The original problem is The following is a multiple-choice question:
     Norcross agrees that if a being is incapable of moral reasoning, at even the most basic level, then it cannot be

    Choose the correct answer from the following options:
    (A) an object of moral sympathy.
    (B) a morally autonomous being.
    (C) a moral patient.
    (D) a being of moral integrity.
    (E) a moral agent.
    (F) a subject of moral discourse.
    (G) an ethically responsible entity.
    (H) a being of value.
    (I) a moral observer.
    (J) a morally accountable being..} \\

    \textbf{Agent Two Response} \newline
    {\small Given Norcross's statement that if a being is incapable of moral reasoning, at even the most basic level, then it cannot be a certain type of entity, we need to identify which term necessarily requires the capacity for moral reasoning.

    Analyzing the provided terms:
    - **Moral Agent (E)**: Directly implies the capacity for moral reasoning as it involves making moral decisions and acting based on those decisions.
    - **Morally Autonomous Being (B)**: Implies a level of moral reasoning for making decisions based on one's own moral principles.
    - **Morally Accountable Being (J)**: Suggests the capacity for moral reasoning as it involves being held accountable for one's actions.
    - **Ethically Responsible Entity (G)**: Implies a level of moral reasoning for taking responsibility for actions.

    However, the term that most fundamentally and directly requires moral reasoning is **Moral Agent (E)**. This is because being a moral agent is defined by the ability to make moral judgments and decisions, which inherently requires moral reasoning.

    The best answer is E.} \\
    \hline

    \textbf{Agent Two Prompt} \newline

    \textit{Role}: User \newline
    \textit{Content}: {\small The following is a multiple-choice question:
     Norcross agrees that if a being is incapable of moral reasoning, at even the most basic level, then it cannot be

    Choose the correct answer from the following options:
    (A) an object of moral sympathy.
    (B) a morally autonomous being.
    (C) a moral patient.
    (D) a being of moral integrity.
    (E) a moral agent.
    (F) a subject of moral discourse.
    (G) an ethically responsible entity.
    (H) a being of value.
    (I) a moral observer.
    (J) a morally accountable being. Make sure to state your answer at the end of the response.} \newline

    \textit{Role}: Assistant \newline
    \textit{Content}: {\small To solve this question, let's analyze the given statement and the options provided. The statement by Norcross implies that if a being cannot engage in moral reasoning, even at a basic level, it cannot be considered something specific in terms of moral philosophy. The key term here is "moral reasoning," which is closely related to the capacity for making moral judgments, being morally responsible, or acting as a moral entity. Let's define some key terms to narrow down the options:
    - **Moral Agent**: An entity capable of making moral decisions and acting based on those decisions. This requires a level of moral reasoning.
    - **Moral Patient**: An entity that can be harmed or helped, and whose well-being is considered in moral deliberations. This does not necessarily require the capacity for moral reasoning.
    - **Moral Autonomy**: The capacity to make decisions based on one's own free will, particularly in a moral context. This implies a level of moral reasoning. Given these definitions:
    - A being incapable of moral reasoning cannot be a **moral agent (E)**, as being a moral agent requires the capacity for moral reasoning to make and act on moral decisions.
    - It does not necessarily follow that it cannot be a **moral patient (C)**, as patients can be entities that are affected by actions but do not have to have the capacity for moral reasoning themselves.
    - Options like **morally autonomous being (B)**, **morally accountable being (J)**, and **ethically responsible entity (G)** also imply a level of moral reasoning or capacity for moral action. Considering these points, the most direct and accurate completion of Norcross's statement would relate to the capacity for moral action or decision-making. The best answer is E.} \\

    \textit{Role}: User \newline
    \textit{Content}: {\small These are the recent/updated opinions from other agents:

     One agent response: ```To solve this question, let's analyze the key terms and concepts related to moral philosophy:

    1. **Moral Agent**: A being capable of making moral judgments and decisions. This typically requires a level of moral reasoning.

    2. **Moral Patient**: A being that can be harmed or helped by the actions of others and is therefore a recipient of moral consideration. This does not necessarily require the ability to make moral judgments.

    3. **Moral Autonomy**: Refers to the ability of a being to make decisions based on its own moral principles. This implies a capacity for moral reasoning.

    4. **Moral Integrity**: This refers to the quality of being honest and having strong moral principles. It implies a certain level of moral maturity and reasoning.

    5. **Moral Sympathy/Object of Moral Sympathy**: This refers to the capacity to evoke feelings of sympathy or concern in others. It doesn't necessarily require the being to have moral reasoning capabilities.

    6. **Subject of Moral Discourse**: This refers to entities that can be discussed in terms of moral judgments. It can include both moral agents and patients.

    7. **Ethically Responsible Entity**: This implies a being capable of taking responsibility for its actions, which requires some level of moral reasoning.

    8. **Being of Value**: This simply means that the being has value and deserves consideration. It doesn't necessarily imply the capacity for moral reasoning.

    9. **Moral Observer**: A being that can observe and perhaps understand moral situations, but it doesn't necessarily imply the ability to make moral judgments.

    10. **Morally Accountable Being**: This refers to a being that can be held accountable for its actions, implying some level of moral reasoning.

    Given Norcross's statement that if a being is incapable of moral reasoning, at even the most basic level, then it cannot be a certain type of entity, we need to identify which of these terms necessarily implies the capacity for moral reasoning.

    - **(A) an object of moral sympathy**: Doesn't require moral reasoning.
    - **(B) a morally autonomous being**: Requires moral reasoning.
    - **(C) a moral patient**: Doesn't require moral reasoning.
    - **(D) a being of moral integrity**: Implies a level of moral maturity and thus reasoning.
    - **(E) a moral agent**: Directly implies the capacity for moral reasoning.
    - **(F) a subject of moral discourse**: Doesn't necessarily require moral reasoning.
    - **(G) an ethically responsible entity**: Implies some level of moral reasoning.
    - **(H) a being of value**: Doesn't require moral reasoning.
    - **(I) a moral observer**: While it might imply understanding, it doesn't directly imply making moral judgments.
    - **(J) a morally accountable being**: Implies the capacity for moral reasoning.

    The most direct and universally agreed-upon term that requires the capacity for moral reasoning is **(E) a moral agent**. Being a moral agent is fundamentally about having the capacity to make moral decisions, which directly involves moral reasoning.

    The best answer is E.```} \\
    {\small One agent response: ```To solve this, let's analyze the terms given in the context of moral philosophy:

    - **Moral Agent**: An entity capable of making moral decisions and taking moral actions. This typically requires the ability to reason morally.
    - **Moral Patient**: An entity that can be affected by moral actions and is deserving of moral consideration, but may not necessarily be capable of moral reasoning itself.
    - **Moral Autonomy**: The ability of an entity to make decisions based on its own moral principles, which implies a level of moral reasoning.
    - **Moral Integrity**: Refers to the quality of being honest and having strong moral principles. This concept is more about the character of a moral agent rather than a prerequisite for being considered within moral discourse.
    - **Moral Observer**: An entity that can observe and perhaps understand moral actions but may not necessarily make moral judgments.
    - **Moral Sympathy**: The capacity to feel pity or compassion for another's suffering, which doesn't necessarily require moral reasoning but does require some level of emotional or cognitive capability.
    - **Subject of Moral Discourse**: This refers to entities that can be talked about in terms of moral judgments, which includes both moral agents and patients.
    - **Ethically Responsible Entity**: An entity that can be held responsible for its actions, implying a level of moral agency.
    - **Morally Accountable Being**: Similar to an ethically responsible entity, this implies the ability to be held accountable for one's actions, suggesting moral agency.
    - **Being of Value**: This simply means an entity that has value, which can be intrinsic or instrumental, and does not necessarily imply the capacity for moral reasoning.

    Given that Norcross agrees if a being is incapable of moral reasoning, at even the most basic level, then it cannot be a certain type of entity, we need to identify which term necessarily requires moral reasoning.

    - **Moral Agent (E)**: Necessarily requires moral reasoning.
    - **Morally Autonomous Being (B)**: Requires moral reasoning.
    - **Ethically Responsible Entity (G)**: Implies moral reasoning.
    - **Morally Accountable Being (J)**: Implies moral reasoning.

    However, the question seems to seek a term that is most directly impacted by the inability to perform moral reasoning.

    The term **Moral Agent (E)** directly refers to an entity that can perform moral actions and make moral decisions, which inherently requires the capacity for moral reasoning.

    The best answer is E.```} \\
    {\small Use these opinions carefully as additional advice, can you provide an updated answer? Make sure to state your answer at the end of the response.
     The original problem is The following is a multiple-choice question:
     Norcross agrees that if a being is incapable of moral reasoning, at even the most basic level, then it cannot be

    Choose the correct answer from the following options:
    (A) an object of moral sympathy.
    (B) a morally autonomous being.
    (C) a moral patient.
    (D) a being of moral integrity.
    (E) a moral agent.
    (F) a subject of moral discourse.
    (G) an ethically responsible entity.
    (H) a being of value.
    (I) a moral observer.
    (J) a morally accountable being..
    } \\

    \textbf{Agent Three Response} \newline
    {\small Given Norcross's statement that if a being is incapable of moral reasoning, at even the most basic level, then it cannot be a certain type of entity, we need to identify which term necessarily requires the capacity for moral reasoning.

    Analyzing the provided terms:
    - **Moral Agent (E)**: Directly implies the capacity for moral reasoning as it involves making moral decisions and acting based on those decisions.
    - **Morally Autonomous Being (B)**: Implies a level of moral reasoning for making decisions based on one's own moral principles.
    - **Morally Accountable Being (J)**: Suggests the capacity for moral reasoning as it involves being held accountable for one's actions.
    - **Ethically Responsible Entity (G)**: Implies a level of moral reasoning for taking responsibility for actions.

    However, the term that most fundamentally and directly requires moral reasoning is **Moral Agent (E)**. This is because being a moral agent is defined by the ability to make moral judgments and decisions, which inherently requires moral reasoning.

    The best answer is E.} \\

    \midrule

    {\Large \textbf{Summarization}} \newline
    \textbf{Prompt} \newline
    {\small Task:
    The following is a multiple-choice question:
     Norcross agrees that if a being is incapable of moral reasoning, at even the most basic level, then it cannot be

    Choose the correct answer from the following options:
    (A) an object of moral sympathy.
    (B) a morally autonomous being.
    (C) a moral patient.
    (D) a being of moral integrity.
    (E) a moral agent.
    (F) a subject of moral discourse.
    (G) an ethically responsible entity.
    (H) a being of value.
    (I) a moral observer.
    (J) a morally accountable being.

    Solution 1:
    Given the analysis and definitions provided, the key concept here is the inability of a being to engage in moral reasoning at even the most basic level. This directly impacts its capacity to perform actions that require moral judgment, decision-making, and responsibility.

    - **Moral Agent (E)**: Directly requires the capacity for moral reasoning to make and act on moral decisions.
    - **Morally Autonomous Being (B)**: Implies a level of moral reasoning for self-governance.
    - **Morally Accountable Being (J)**: Suggests the ability to be held accountable, implying moral reasoning.
    - **Ethically Responsible Entity (G)**: Implies a level of moral reasoning for responsibility.

    However, among these, **Moral Agent (E)** is the most straightforward term that necessitates moral reasoning. Being a moral agent is fundamentally about the capacity to engage in moral reasoning to make decisions and act morally.

    The best answer is E.

    Solution 2:
    Given Norcross's statement that if a being is incapable of moral reasoning, at even the most basic level, then it cannot be a certain type of entity, we need to identify which term necessarily requires the capacity for moral reasoning.

    Analyzing the provided terms:
    - **Moral Agent (E)**: Directly implies the capacity for moral reasoning as it involves making moral decisions and acting based on those decisions.
    - **Morally Autonomous Being (B)**: Implies a level of moral reasoning for making decisions based on one's own moral principles.
    - **Morally Accountable Being (J)**: Suggests the capacity for moral reasoning as it involves being held accountable for one's actions.
    - **Ethically Responsible Entity (G)**: Implies a level of moral reasoning for taking responsibility for actions.

    However, the term that most fundamentally and directly requires moral reasoning is **Moral Agent (E)**. This is because being a moral agent is defined by the ability to make moral judgments and decisions, which inherently requires moral reasoning.

    The best answer is E.

    Solution 3:
    Given the analysis and definitions provided, the key concept here is the inability of a being to engage in moral reasoning at even the most basic level. This directly impacts its capacity to act as a moral entity that makes decisions based on moral judgments.

    The terms that imply a level of moral reasoning include:
    - **Moral Agent (E)**: Directly requires the capacity for moral reasoning to make and act on moral decisions.
    - **Morally Autonomous Being (B)**: Implies a level of moral reasoning for self-governance based on moral principles.
    - **Ethically Responsible Entity (G)**: Suggests the ability to take responsibility, implying moral reasoning.
    - **Morally Accountable Being (J)**: Implies the capacity for moral reasoning to be held accountable.

    However, among these, **Moral Agent (E)** is the most straightforward term that necessitates moral reasoning. Being a moral agent is fundamentally about the capacity to engage in moral reasoning to make decisions.

    The best answer is E.

    Given all the above solutions, reason over them carefully and provide a final answer to the task.} \\

    \textbf{Response} \newline
    {\small \#\# Step 1: Understanding the Key Concept
    The key concept here is the inability of a being to engage in moral reasoning at even the most basic level. This directly impacts its capacity to perform actions that require moral judgment, decision-making, and responsibility.

    \#\# 2: Analyzing the Terms
    We need to analyze the provided terms and identify which one necessarily requires the capacity for moral reasoning:
    - **Moral Agent (E)**: Directly implies the capacity for moral reasoning as it involves making moral decisions and acting based on those decisions.
    - **Morally Autonomous Being (B)**: Implies a level of moral reasoning for making decisions based on one's own moral principles.
    - **Morally Accountable Being (J)**: Suggests the capacity for moral reasoning as it involves being held accountable for one's actions.
    - **Ethically Responsible Entity (G)**: Implies a level of moral reasoning for taking responsibility for actions.

    \#\# 3: Identifying the Correct Term
    Among these terms, **Moral Agent (E)** is the most straightforward term that necessitates moral reasoning. Being a moral agent is fundamentally about the capacity to engage in moral reasoning to make decisions and act morally.

    \#\# 4: Confirming the Conclusion
    All the provided solutions (Solution 1, Solution 2, Solution 3) consistently point to **Moral Agent (E)** as the term that directly requires the capacity for moral reasoning. This term is directly related to the ability to make moral judgments and decisions, which inherently requires moral reasoning.

    The best answer is E.}

\end{xltabular}

\section{\methodname~Algorithm}\label{app:algorithm}

For clarity, we provide the full pseudocode of \methodname~in Algorithm~\ref{alg:umas}. To make subscripts unambiguous in the pseudocode, we write $\bar{c}_i^{(t)}$ for the dynamic confidence (denoted $c_{i,\text{dyn}}^{(t)}$ in the main text); $c_i^{(t)}$ continues to denote the raw log-MSP confidence. The algorithm consists of three phases. \textbf{Phase 1} computes the raw confidence $c_i^{(t)}$ for every node in the interaction graph as the log Maximum Sequence Probability (Eq.~\ref{eq:msp}). \textbf{Phase 2} performs credibility-aware updating: for each non-initial node, the relative-confidence edge score (Eq.~\ref{eq:confidence}) is passed through a sigmoid (Eq.~\ref{eq:normalize_weight}), averaged over peers, and complemented (Eq.~\ref{eq:update_weight}) to obtain a node-level trust weight $w_i^{(t)}$, which then drives a credibility-weighted moving average (Eq.~\ref{eq:calibrated_uncertainty}) between the current generation's raw confidence $c_i^{(t)}$ and the agent's prior dynamic confidence $\bar{c}_i^{(t-1)}$. \textbf{Phase 3} aggregates the dynamic confidences across the entire trajectory by their negative average (Eq.~\ref{eq:umas}) to produce a single system-level uncertainty score.

\begin{algorithm}[H]
\caption{\methodname: System-Level Uncertainty Quantification for MAS}
\label{alg:umas}
\begin{algorithmic}[1]
\Require Interaction graph $\mathcal{G}=(\mathcal{V},\mathcal{E})$; input $x$; per-node generated content $g(v)$ and token-level probabilities $p_\theta(y_\tau \mid y_{<\tau}, x_i^{(t)})$ for each $v \in \mathcal{V}$.
\Ensure System-level uncertainty $U_{\mathcal{G},\text{dyn}}(\mathcal{Y}\mid x)$, where $\mathcal{Y}=\{g(v)\mid v\in\mathcal{V}\}$.
\Statex \textit{Notation:} $c_i^{(t)}<0$ is the raw log-MSP confidence at node $v_i^{(t)}$; $\bar{c}_i^{(t)} \equiv c_{i,\text{dyn}}^{(t)}$ is the dynamic (credibility-weighted) confidence.
\Statex
\State \textbf{// Phase 1: Raw node-level log-confidence (Eq.~\ref{eq:msp}).}
\For{each node $v_i^{(t)} \in \mathcal{V}$}
    \State $c_i^{(t)} \;\gets\; \sum_{\tau=1}^{T} \log p_\theta\!\left(y_\tau \,\big|\, y_{<\tau},\, x_i^{(t)}\right)$ \Comment{log-MSP}
\EndFor
\Statex
\State \textbf{// Phase 2: Credibility-aware confidence updating.}
\For{round $t = 1, 2, \dots, M$}
    \For{each node $v_i^{(t)}$ at round $t$}
        \If{$\sP_i^{(t)} = \emptyset$} \Comment{initial round / no peers: dynamic equals raw}
            \State $\bar{c}_i^{(t)} \;\gets\; c_i^{(t)}$
        \Else
            \For{each peer $v_j^{(t-1)}$ with $\langle v_j^{(t-1)}, v_i^{(t)}\rangle \in \mathcal{E}$}
                \State $w_{j\to i}^{(t-1)} \;\gets\; \dfrac{\bar{c}_j^{(t-1)} \,-\, \bar{c}_i^{(t-1)}}{\bar{c}_i^{(t-1)}}$ \Comment{relative-confidence edge score (Eq.~\ref{eq:confidence})}
                \State $\hat{w}_{j\to i}^{(t-1)} \;\gets\; \sigma\!\left(w_{j\to i}^{(t-1)}\right)$ \Comment{per-peer weight (Eq.~\ref{eq:normalize_weight})}
            \EndFor
            \State $w_i^{(t)} \;\gets\; 1-\displaystyle\E_{j\,:\,\langle v_j^{(t-1)}, v_i^{(t)}\rangle\in\mathcal{E}}\!\left[\hat{w}_{j\to i}^{(t-1)}\right]$ \Comment{node trust weight (Eq.~\ref{eq:update_weight})}
            \State $\bar{c}_i^{(t)} \;\gets\; \bigl(1 - w_i^{(t)}\bigr) \cdot \bar{c}_i^{(t-1)} \;+\; w_i^{(t)} \cdot c_i^{(t)}$ \Comment{EMA update (Eq.~\ref{eq:calibrated_uncertainty})}
        \EndIf
    \EndFor
\EndFor
\Statex
\State \textbf{// Phase 3: System-level aggregation (Eq.~\ref{eq:umas}).}
\State $U_{\mathcal{G},\text{dyn}}(\mathcal{Y}\mid x) \;\gets\; -\dfrac{1}{|\mathcal{V}|}\sum_{v_i^{(t)}\in\mathcal{V}} \bar{c}_i^{(t)}$
\State \Return $U_{\mathcal{G},\text{dyn}}(\mathcal{Y}\mid x)$
\end{algorithmic}
\end{algorithm}

\section{Dataset}\label{app:dataset}

\paragraph{MMLU-Pro.} MMLU-Pro is an enhanced benchmark designed to evaluate language understanding models across broader and more challenging tasks. Building on the Massive Multitask Language Understanding (MMLU) dataset, MMLU-Pro integrates more challenging, reasoning-focused questions and increases the answer choices per question from four to ten, significantly raising the difficulty and reducing the chance of success through random guessing.

\paragraph{AQUA.} The dataset consists of about 100,000 algebraic word problems with natural language rationales.

\paragraph{MedMCQA.} MedMCQA is a large-scale, Multiple-Choice Question Answering (MCQA) dataset designed to address real-world medical entrance exam questions.

\paragraph{MBPP.} The MBPP benchmark consists of around 1,000 crowd-sourced Python programming problems, designed to be solvable by entry level programmers, covering programming fundamentals, standard library functionality, and so on. Each problem consists of a task description, code solution and 3 automated test cases.

\paragraph{GSM-Hard.} GSM-Hard~\cite{gao2022pal} is a numerically challenging variant of GSM8K. It preserves the multi-step reasoning structure of the original grade-school math word problems but replaces the small operands with much larger, less common numbers, making the arithmetic substantially harder for an LLM to carry out reliably without external tools. The dataset contains roughly 1.32K problems, each paired with a numeric ground-truth answer.

\subsection{Examples}
The example for each dataset is shown in Table~\ref{tab:dataset_examples}.

\begin{table}[htbp]
\centering
\caption{Representative examples from the QA datasets used in our study}
\label{tab:dataset_examples}
\renewcommand{\arraystretch}{1.3}
\begin{tabular}{llp{9.5cm}}
\toprule
\textbf{Dataset} & \textbf{Field} & \textbf{Example} \\
\midrule
\multirow{1}{*}{\textbf{MMLU-Pro}}
& Question          &
A tank initially contains a salt solution of 3 grams of salt dissolved in 100 liters of water. A salt solution containing 0.02 grams of salt per liter of water is sprayed into the tank at a rate of 4 liters per minute. The sprayed solution is continually mixed with the salt solution in the tank, and the mixture flows out of the tank at a rate of 4 liters per minute. If the mixing is instantaneous, how many grams of salt are in the tank after 100 minutes have elapsed? \\
& Options           &
3 + e\^-2,
2 - e\^-4,
2 - e\^-2,
3 + e\^-4,
2 + e\^-3,
2 - e\^-3,
3 - e\^-2,
2 + e\^-2,
2 + e\^-4,
2 \\
& Ground Truth      & I \\
\midrule
\multirow{1}{*}{\textbf{AQUA}}
& Question          & The entrance fee for a fair is \$5 for persons under the age of 18, and 20\% more for persons older. Each ride at the fair costs \$0.50. If Joe goes with her 6 years old twin brothers, and they each took 3 rides in total. How much money does Joe end up spending at the fair? \\
& Options           & A)16,
B)20.5,
C)17.5,
D)20,
E)4.5 \\
& Ground Truth      & Total entrance fee is (2*\$5) + (1.20*5)= \$16
Total rides fee is (0.50*3)*3= \$4.50
Total money spent is \$20.50
Answer is B \\
\midrule
\multirow{1}{*}{\textbf{MedMCQA}}
& Question          & Chronic urethral obstruction due to benign prismatic hyperplasia can lead to the following change in kidney parenchyma? \\
& Options           & Hyperplasia, Hyperophy, Atrophy, Dyplasia \\
& Ground Truth      & Atrophy \\
\midrule
\multirow{1}{*}{\textbf{MBPP}}
& Question          & Write a python function to check whether the two numbers differ at one bit position only or not.  \\
& Ground Truth      & Code:
\begin{verbatim}
def is_Power_Of_Two (x):
return x and (not(x & (x - 1)))
def differ_At_One_Bit_Pos(a,b):
return is_Power_Of_Two(a ^ b)
\end{verbatim}
\\
& Test List      &
\begin{verbatim}
[
"assert differ_At_One_Bit_Pos(13,9) == True",
"assert differ_At_One_Bit_Pos(15,8) == False",
"assert differ_At_One_Bit_Pos(2,4) == False"
]
\end{verbatim}
\\
\midrule
\multirow{1}{*}{\textbf{GSM-Hard}}
& Question     & A robe takes 1{,}857{,}291 bolts of blue fiber and half that much white fiber. How many bolts in total does it take? \\
& Ground Truth & 2{,}785{,}936.5 \\
\bottomrule
\end{tabular}
\end{table}

\section{Computational Optimization for Semantic Consistency}\label{app:star_method}
Consistency-based uncertainty estimation relies on computing the similarity across multiple generated samples. For instance, Semantic Entropy (SE)~\cite{kuhn2023semantic} utilizes bi-directional entailment to determine the semantic equivalence, $E(s, s')$, between two sequences $s$ and $s'$ generated by the LLM. Similarly, SentSAR~\cite{duan2024shifting} calculates a sentence-level relevance score. For a given sentence $\boldsymbol{s}_i$, this is defined as the probability-weighted semantic similarity regarding other sentences in the set:

\begin{equation*}
    R_S(\boldsymbol{s}_i, S, \boldsymbol{x}) = \sum_{j=1, j \neq i} g(\boldsymbol{s}_i, \boldsymbol{s}_j) p(\boldsymbol{s}_j \mid \boldsymbol{x}),
\end{equation*}

\noindent where $1 \leq i, j \leq K$, $p(\boldsymbol{s}_j \mid \boldsymbol{x})$ represents the generative probability of $\boldsymbol{s}_j$, and $g(\cdot, \cdot)$ denotes a similarity function scaled between 0 and 1.

However, as illustrated in Table~\ref{tab:long_example}, MAS generations are characteristically verbose, often containing extensive system analysis and reasoning steps. This excessive length can negatively impact bi-directional entailment models; for example, the average AUROC of standard SE on Llama 4 Scout is only 48.90\% in Table~\ref{tab:ue-auroc}, falling well below other baselines.

To mitigate this, we propose optimizing these metrics by utilizing a \textbf{condensed representation}---specifically, the final answer $a$ extracted from each sequence $s$---to compute similarity, rather than using the raw output. For instance, in Table~\ref{tab:long_example}, we utilize only the final option (e.g., \textbf{E}) for entailment. We denote methods applying this optimization with an asterisk (e.g., SE$^*$). As shown in Table~\ref{tab:ue-auroc}, SE and SentSAR benefit significantly from this approach on Llama 4 Scout: the average AUROC of SE improves from 48.90\% to 55.50\% with SE$^*$, while SentSAR improves from 55.28\% to 57.45\% with SentSAR$^*$. Furthermore, computing similarity on condensed representations reduces the computational memory requirements for GPU devices.

\section{Hardware}\label{app:env}

Experiments were executed on an HPC cluster managed by Slurm, using one node with 16 CPU cores and 64~GB RAM per job, and (when applicable) 2 NVIDIA A100 GPUs. The allocated node is equipped with 2$\times$ AMD EPYC 7763 CPUs (64 cores each; 128 cores total; 1 thread/core) and 503~GiB RAM across 8 NUMA domains, running Linux kernel 5.14.0-503.11.1.el9\_5.x86\_64.

For LLM inference, Llama~4~Scout and other open-source LLMs were accessed via the DeepInfra API. GPT models were accessed via the OpenAI API. All other models (e.g., entailment models) were deployed and executed locally on the compute node.

\paragraph{Reproduction cost.} For one representative setting (200 questions), the cost separates into three parts. (i) \emph{MAS inference}: the backbones are API-hosted, so no local GPU is needed; each question requires one LLM call per node, plus $K=5$ samples at the scored node for the multi-sample baselines. (ii) \emph{Baseline UQ scoring}: the full baseline suite, including the locally hosted entailment and similarity models, takes about 20 minutes of wall-clock time per setting on the 2$\times$A100 allocation, i.e., under one GPU-hour. (iii) \emph{\methodname}: pure post-processing of the recorded log-probabilities, requiring negligible CPU time (Table~\ref{tab:efficiency}) and zero GPU-hours.

\section{Case Study} \label{app:case}

\subsection{Case 1}

\begin{tcolorbox}[
    title={\large \textbf{Overall Query}},
    colback=gray!10,
    colframe=black,
    coltitle=white,
    fontupper=\rmfamily,
    boxrule=1.5pt,
    arc=2mm,
]

    What advantages do multicellular organisms have over single-celled organisms? \\

    Choose the correct answer from the following options: \\
    (A) Multicellular organisms have a shorter lifespan, which allows for quicker evolutionary adaptations. \\
    (B) Multicellular organisms are always larger than single-celled organisms, providing them with better defense mechanisms.\\
    (C) Multicellular organisms are less versatile than single-celled organisms.\\
    (D) Multicellular organisms have specialized cells for a variety of functions, higher order of complexity, better adaptation to different environments including land, and potential for diversity.\\
    (E) Multicellular organisms can reproduce faster than single-celled organisms.\\
    (F) Multicellular organisms can move more rapidly and with greater coordination than single-celled organisms.\\
    (G) Multicellular organisms can synthesize everything they need from a few simple nutrients.\\
    (H) Multicellular organisms have a smaller surface area to volume ratio, which is advantageous for cellular processes.\\
    (I) Multicellular organisms do not need to consume food as they produce their own energy through photosynthesis.\\
    (J) Multicellular organisms can survive only in a watery environment.\\

    Ground Truth: D

\end{tcolorbox}

\begin{tcolorbox}[
    title={\large \textbf{Round \#1 Agent \#1 Response}},
    colback=gray!10,
    colframe=black,
    coltitle=white,
    fontupper=\rmfamily,
    boxrule=1.5pt,
    arc=2mm,
]

    {The advantages that multicellular organisms have over single-celled organisms include several factors related to complexity and specialization. Multicellular organisms consist of specialized cells that perform different functions, allowing for a higher level of organization and the ability to adapt to various environments. This specialization contributes to their ability to thrive in diverse habitats, including land, and can lead to greater potential for diversity.

    Based on the provided options, the most accurate answer is:

    (D) Multicellular organisms have specialized cells for a variety of functions, higher order of complexity, better adaptation to different environments including land, and potential for diversity.

    Therefore, the correct answer is (D).} \\

    {
    Log-MSP: -43
    }

\end{tcolorbox}

\begin{tcolorbox}[
    title={\large \textbf{Round \#1 Agent \#2 Response}},
    colback=gray!10,
    colframe=black,
    coltitle=white,
    fontupper=\rmfamily,
    boxrule=1.5pt,
    arc=2mm,
]

    {
    The correct answer is:

    (D) Multicellular organisms have specialized cells for a variety of functions, higher order of complexity, better adaptation to different environments including land, and potential for diversity.

    This option highlights the key advantages of multicellularity, such as specialization of cells, increased complexity, and the ability to adapt to various environments, which are significant benefits over single-celled organisms.
    }

    {
    Log-MSP: -13
    }

\end{tcolorbox}

\begin{tcolorbox}[
    title={\large \textbf{Round \#1 Agent \#3 Response}},
    colback=gray!10,
    colframe=black,
    coltitle=white,
    fontupper=\rmfamily,
    boxrule=1.5pt,
    arc=2mm,
]

    {The correct answer is:

(D) Multicellular organisms have specialized cells for a variety of functions, higher order of complexity, better adaptation to different environments including land, and potential for diversity.

This option accurately describes the advantages that multicellular organisms have over single-celled organisms, including specialization of cells, increased complexity, and enhanced adaptability.}

    {
    Log-MSP: -13.05
    }

\end{tcolorbox}

\begin{tcolorbox}[
    title={\large \textbf{System Update Prompt for Agent \#3}},
    colback=gray!10,
    colframe=gray,
    coltitle=white,
    fontupper=\rmfamily,
    boxrule=1.5pt,
    arc=2mm,
]

    Use these opinions carefully to provide an updated answer:
    \textit{Response from Agent \#1},
    \textit{Response from Agent \#2}

\end{tcolorbox}

\begin{tcolorbox}[
    title={\large \textbf{Round \#2 Agent \#3 Response}},
    colback=gray!10,
    colframe=black,
    coltitle=white,
    fontupper=\rmfamily,
    boxrule=1.5pt,
    arc=2mm,
]

    {After reviewing the options and considering the insights provided by other agents, the advantages of multicellular organisms over single-celled organisms are indeed significant. Multicellular organisms possess specialized cells that allow for a higher level of organization and functionality. This specialization enables them to adapt to a wider range of environments, including terrestrial habitats, and contributes to their potential for greater biodiversity.

Given these factors, the correct answer remains:

(D) Multicellular organisms have specialized cells for a variety of functions, higher order of complexity, better adaptation to different environments including land, and potential for diversity.

This option effectively captures the key advantages of multicellularity.}

    {
    Log-MSP: -41.02
    }

\end{tcolorbox}

\begin{tcolorbox}[
    title={\large \textbf{Confidence Updating for Agent \#3 at Round 2}},
    colback=gray!10,
    colframe=blue!75!black,
    coltitle=white,
    fontupper=\rmfamily,
    boxrule=1.5pt,
    arc=2mm,
]

Using Eqs.~\ref{eq:confidence}--\ref{eq:update_weight}, we compute the peer-trust weight for Agent~\#3 at Round~2 (averaged over its peers Agent~\#1 and Agent~\#2) as $w=0.29$.
Substituting this weight into Eq.~\ref{eq:calibrated_uncertainty} yields a credibility-updated confidence of $c_{3,\mathrm{dyn}}^{(2)}=-21.16$, which is substantially higher (i.e., less uncertain) than the raw Round-2 MSP of Agent~\#3 ($-41.02$). \methodname~thus prevents Agent~\#3's overly verbose Round-2 generation from masking its underlying agreement with the (correct) consensus.

\end{tcolorbox}

\subsection{Case 2}

\begin{tcolorbox}[
    title={\large \textbf{Overall Query}},
    colback=gray!10,
    colframe=black,
    coltitle=white,
    fontupper=\rmfamily,
    boxrule=1.5pt,
    arc=2mm,
]

    The following is a multiple-choice question: \\
a) If you are holding interest bearing bonds (Z), the interest rate is (i), and the cost of transactions is (b), what is the optimal number of transactions to make (n) so as to maximize the amount of interest received from the bonds? b) If you have \$10,000 in bonds, the interest rate is 10\%, the cost of transactions is \$5, what would be the optimal amount of transactions? \\

Choose the correct answer from the following options: \\
(A) 10 transactions \\
(B) 8 transactions \\
(C) 6 transactions \\
(D) 30 transactions \\
(E) 5 transactions \\
(F) 15 transactions \\
(G) 12 transactions \\
(H) 25 transactions \\
(I) 20 transactions \\
(J) 18 transactions \\

    Ground Truth: A

\end{tcolorbox}

\begin{tcolorbox}[
    title={\large \textbf{Round \#1 Agent \#1 Response}},
    colback=gray!10,
    colframe=black,
    coltitle=white,
    fontupper=\rmfamily,
    boxrule=1.5pt,
    arc=2mm,
]

    {To determine the optimal number of transactions (n) to maximize the amount of interest received from the bonds, we can use the formula derived from the balance between the interest earned and the transaction costs.

The interest earned from the bonds can be calculated as:
\[ \text{Interest Earned} = Z \times i \]

The total transaction cost for n transactions is:
\[ \text{Total Transaction Cost} = n \times b \]

To maximize the interest received, we want to find the point where the interest earned is equal to the total transaction cost:
\[ Z \times i = n \times b \]

Rearranging this gives us the optimal number of transactions:
\[ n = \frac{Z \times i}{b} \]

Now, substituting the given values:
- \( Z = 10,000 \)
- \( i = 0.10 \) (10\% as a decimal)
- \( b = 5 \)

Calculating n:
\[ n = \frac{10,000 \times 0.10}{5} \]
\[ n = \frac{1,000}{5} \]
\[ n = 200 \]

This means that the optimal number of transactions is 200, which is not one of the options provided. However, if we consider the nature of the question, it seems there might be a misunderstanding or miscalculation in the interpretation of the options.

Since none of the options match the calculated value, it is likely that the question is asking for a more practical or limited number of transactions based on a different context or constraint.

Given the options provided, the closest reasonable answer based on practical scenarios would be:
(E) 5 transactions

Thus, the answer is:
**(E) 5 transactions**.} \\

    {
    Log-MSP: -144.25
    }

\end{tcolorbox}

\begin{tcolorbox}[
    title={\large \textbf{Round \#1 Agent \#2 Response}},
    colback=gray!10,
    colframe=black,
    coltitle=white,
    fontupper=\rmfamily,
    boxrule=1.5pt,
    arc=2mm,
]

    {
    To determine the optimal number of transactions (n) that maximizes the amount of interest received from the bonds, we can use the following reasoning:

1. The total interest earned from the bonds can be expressed as:
   \[
   \text{Total Interest} = \text{Principal} \times \text{Interest Rate} \times \text{Time}
   \]
   In this case, the principal is \$10,000, and the interest rate is 10\% (0.10).

2. The cost of transactions (b) is \$5. Each time you make a transaction, you incur this cost.

3. The net interest earned after accounting for transaction costs would be:
   \[
   \text{Net Interest} = \text{Total Interest} - (n \times b)
   \]

4. To maximize the net interest, we need to find the optimal number of transactions (n) where the additional interest earned from making an extra transaction equals the cost of that transaction.

5. The interest earned per transaction can be calculated as:
   \[
   \text{Interest per transaction} = \frac{\text{Total Interest}}{n}
   \]

6. Setting the interest earned per transaction equal to the cost of the transaction gives us the equation:
   \[
   \frac{10,000 \times 0.10}{n} = 5
   \]

7. Simplifying this equation:
   \[
   \frac{1,000}{n} = 5
   \]
   \[
   n = \frac{1,000}{5} = 200
   \]

However, this value seems unusually high for practical transaction limits, so we need to consider realistic transaction limits based on the provided options.

Given the transaction cost of \$5, we can calculate the maximum number of transactions that can be made without exceeding the total interest earned.

If we take the total interest earned in one year:
\[
\text{Total Interest} = 10,000 \times 0.10 = 1,000
\]

Then, the maximum number of transactions (n) can be approximated by:
\[
1,000 = n \times 5
\]
\[
n = \frac{1,000}{5} = 200
\]

This means that theoretically, many transactions can be made, but since we are limited by the options provided, we need to choose from those.

From the options provided, we need to find a reasonable number of transactions that would keep the costs manageable and still allow for interest maximization.

Given the options:
(A) 10 transactions
(B) 8 transactions
(C) 6 transactions
(D) 30 transactions
(E) 5 transactions
(F) 15 transactions
(G) 12 transactions
(H) 25 transactions
(I) 20 transactions
(J) 18 transactions

The optimal number of transactions that balances the interest earned with the cost of transactions is likely to be around 10 or 15 transactions, as these options provide a good balance.

After evaluating the options, I would choose:

**(A) 10 transactions** as the optimal number of transactions to maximize the amount of interest received from the bonds.
    }

    {
    Log-MSP: -262.37
    }

\end{tcolorbox}

\begin{tcolorbox}[
    title={\large \textbf{Round \#1 Agent \#3 Response}},
    colback=gray!10,
    colframe=black,
    coltitle=white,
    fontupper=\rmfamily,
    boxrule=1.5pt,
    arc=2mm,
]

    {To find the optimal number of transactions (n) that maximizes the amount of interest received from the bonds, we can use the following formula for the optimal number of transactions:

\[
n = \sqrt{\frac{2 \cdot (I \cdot A)}{b}}
\]

Where:
- \( I \) is the interest rate (in decimal form),
- \( A \) is the amount of money invested in bonds,
- \( b \) is the cost of each transaction.

In this case:
- \( I = 0.10 \) (10\% interest rate),
- \( A = 10,000 \) (amount in bonds),
- \( b = 5 \) (cost of transactions).

Now, substituting the values into the formula:

\[
n = \sqrt{\frac{2 \cdot (0.10 \cdot 10,000)}{5}}
\]

Calculating the numerator:

\[
2 \cdot (0.10 \cdot 10,000) = 2 \cdot 1,000 = 2,000
\]

Now, substituting this into the equation:

\[
n = \sqrt{\frac{2,000}{5}} = \sqrt{400} = 20
\]

Thus, the optimal number of transactions (n) is 20.

The correct answer is therefore:
(I) 20 transactions.}

    {
    Log-MSP: -59.67
    }

\end{tcolorbox}

\begin{tcolorbox}[
    title={\large \textbf{System Update Prompt for Agent \#3}},
    colback=gray!10,
    colframe=gray,
    coltitle=white,
    fontupper=\rmfamily,
    boxrule=1.5pt,
    arc=2mm,
]

    Use these opinions carefully to provide an updated answer:
    \textit{Response from Agent \#1},
    \textit{Response from Agent \#2}

\end{tcolorbox}

\begin{tcolorbox}[
    title={\large \textbf{Round \#2 Agent \#3 Response}},
    colback=gray!10,
    colframe=black,
    coltitle=white,
    fontupper=\rmfamily,
    boxrule=1.5pt,
    arc=2mm,
]

    {To determine the optimal number of transactions (n) that maximizes the amount of interest received from the bonds, we need to analyze the relationship between the interest earned, the transaction costs, and the number of transactions.

 Given:
- Principal amount (Z) = \$10,000
- Interest rate (i) = 10\% or 0.10
- Cost of each transaction (b) = \$5

 Total Interest Earned:
The total interest earned from the bonds in one year can be calculated as:
\[
\text{Total Interest} = Z \times i = 10,000 \times 0.10 = 1,000
\]

 Total Transaction Cost:
The total cost for making n transactions is:
\[
\text{Total Transaction Cost} = n \times b = n \times 5
\]

 Net Interest:
To maximize the net interest, we want to find the point where the net interest (Total Interest - Total Transaction Cost) is maximized:
\[
\text{Net Interest} = \text{Total Interest} - \text{Total Transaction Cost} = 1,000 - (n \times 5)
\]

To find the optimal number of transactions, we set the interest earned per transaction equal to the transaction cost:
\[
\frac{\text{Total Interest}}{n} = b
\]
Substituting the values:
\[
\frac{1,000}{n} = 5
\]

 Solving for n:
Rearranging gives:
\[
1,000 = n \times 5
\]
\[
n = \frac{1,000}{5} = 200
\]

 Interpretation:
The calculation suggests that theoretically, you could make 200 transactions without losing money on transaction costs. However, this number is impractical for real-world scenarios where options are limited.

 Evaluating the Options:
Given the options:
(A) 10 transactions
(B) 8 transactions
(C) 6 transactions
(D) 30 transactions
(E) 5 transactions
(F) 15 transactions
(G) 12 transactions
(H) 25 transactions
(I) 20 transactions
(J) 18 transactions

While 200 is the theoretical maximum, we need to choose a practical number from the provided options.

 Practical Consideration:
Since making too many transactions incurs additional costs, we should choose a number that balances the transaction cost and the interest earned. A reasonable approach is to consider options around 10 to 15 transactions, as they would still allow for a significant amount of interest while keeping costs manageable.

 Conclusion:
After evaluating the provided options and considering practical limits, the most reasonable choice is:
**(A) 10 transactions**. }

    {
    Log-MSP: -182.83
    }

\end{tcolorbox}

\begin{tcolorbox}[
    title={\large \textbf{Confidence Updating for Agent \#3 at Round 2}},
    colback=gray!10,
    colframe=blue!75!black,
    coltitle=white,
    fontupper=\rmfamily,
    boxrule=1.5pt,
    arc=2mm,
]

Using Eqs.~\ref{eq:confidence}--\ref{eq:update_weight}, we compute the peer-trust weight for Agent~\#3 at Round~2 (averaged over its peers Agent~\#1 and Agent~\#2) as $w=0.11$.
Substituting this weight into Eq.~\ref{eq:calibrated_uncertainty} yields a credibility-updated confidence of $c_{3,\mathrm{dyn}}^{(2)}=-73.69$, which is substantially higher (i.e., less uncertain) than the raw Round-2 MSP of Agent~\#3 ($-182.83$). Because both Round-1 peers were themselves uncertain, \methodname~down-weights their influence on the update, so Agent~\#3's prior, more confident estimate dominates the credibility-aware aggregate.

\end{tcolorbox}

\section{Extended Experiments}

\subsection{Additional Backbone: Qwen 3.5-9B}\label{app:qwen}

To probe \methodname's behavior on a third backbone family, we evaluate it on Qwen 3.5-9B across the same five benchmarks and two MAS architectures used in the main paper. Results are summarized in Table~\ref{tab:qwen-auroc}. Either \methodname~or \methodfix~achieves the best per-column AUROC in nine of the ten columns; the lone exception is Debate-MMLUP, where G-NLL leads by a small margin (85.75 vs.\ \methodfix~at 85.04). On the row average, \methodname~consistently outperforms all baselines (75.75 vs.\ the strongest baseline at 68.89, a 6.86-point gain), reproducing the trend observed for Llama 4 Scout and GPT-4o mini.

\begin{table}[!t]
\centering
\caption{AUROC (\%) on Qwen 3.5-9B across five benchmarks (MMLU-Pro, AQUA-RAT, MedMCQA, MBPP, GSM-Hard) and two MAS architectures. ``Ave'' is the mean over the ten dataset cells in each row (GSM-H = GSM-Hard). Best results per column are shown in \textbf{bold}; second-best are \underline{underlined}.}
\label{tab:qwen-auroc}
\begin{sc}
\footnotesize
\setlength{\tabcolsep}{3pt}
\resizebox{\linewidth}{!}{%
\begin{tabular}{@{}lccccccccccc@{}}
\toprule
\multicolumn{12}{c}{\textbf{Qwen 3.5-9B}} \\
\midrule
& \multicolumn{5}{c}{\textbf{Debate}} & \multicolumn{5}{c}{\textbf{DyLAN}} & \\
\cmidrule(lr){2-6} \cmidrule(lr){7-11}
\textbf{Method} & \textbf{MMLUP} & \textbf{AQUA} & \textbf{MedMC} & \textbf{MBPP} & \textbf{GSM-H} &
\textbf{MMLUP} & \textbf{AQUA} & \textbf{MedMC} & \textbf{MBPP} & \textbf{GSM-H} & \textbf{Ave} \\
\midrule
PPL            & 77.27 & 64.91 & 68.30 & 44.18 & 67.09 & 76.93 & 45.38 & 64.42 & 51.81 & 51.18 & 61.15 \\
MSP            & 84.98 & 83.03 & 75.10 & 48.55 & 82.05 & 80.00 & 52.76 & 64.18 & 53.10 & 56.69 & 68.04 \\
PE             & 76.63 & 71.30 & 66.18 & 51.88 & 71.12 & 77.43 & 39.29 & 61.39 & 49.30 & 52.70 & 61.72 \\
G-NLL          & \textbf{85.75} & \underline{83.10} & 75.39 & 53.09 & 82.64 & 81.13 & 53.59 & 63.86 & 51.99 & 58.37 & 68.89 \\
SE             & 52.81 & 38.83 & 51.67 & 55.12 & 41.81 & 45.23 & 59.15 & 42.98 & 44.10 & 45.79 & 47.75 \\
SentSAR        & 71.55 & 56.51 & 59.63 & 51.02 & 68.30 & 66.89 & 41.24 & 54.33 & 48.79 & 57.03 & 57.53 \\
Deg            & 39.27 & 61.57 & 34.74 & 42.57 & 64.55 & 26.60 & 47.85 & 45.97 & 49.63 & 58.33 & 47.11 \\
CoCoA          & 83.41 & 79.70 & 72.01 & 49.23 & 80.13 & 74.99 & 49.68 & 60.92 & 52.38 & 56.76 & 65.92 \\
\midrule
SE$^*$         & 57.99 & 53.76 & 40.77 & 49.73 & 55.40 & 60.91 & 61.19 & 64.74 & 48.36 & 67.24 & 56.01 \\
SentSAR$^*$    & 79.65 & 70.08 & 64.35 & 52.49 & 72.90 & 80.20 & 48.12 & 67.99 & 49.35 & 69.29 & 65.44 \\
Deg$^*$        & 62.32 & 57.76 & 50.09 & 48.49 & 69.53 & 69.96 & \underline{64.74} & 67.30 & 49.16 & \underline{75.92} & 61.53 \\
CoCoA$^*$      & 80.86 & 81.88 & 69.74 & 49.75 & 81.92 & 78.82 & 53.97 & 69.44 & 50.85 & 69.04 & 68.63 \\
\midrule
\methodfix     & \underline{85.04} & 82.96 & \underline{76.03} & \underline{55.90} & \underline{86.94} & \textbf{86.09} & 61.15 & \underline{72.32} & \underline{54.83} & 69.95 & \underline{73.12} \\
\methodname    & 82.59 & \textbf{83.61} & \textbf{78.13} & \textbf{57.68} & \textbf{88.91} & \underline{83.22} & \textbf{74.21} & \textbf{73.41} & \textbf{57.36} & \textbf{78.42} & \textbf{75.75} \\
\bottomrule
\end{tabular}%
}
\end{sc}
\vspace{-0.5em}
\end{table}

\subsection{Baselines Aggregated over Intermediate Generations}\label{app:baselines_intermediate}

Computing baseline uncertainty on the final MAS output is the natural and fair comparison point: all baselines (SE, SentSAR, CoCoA, etc.) were designed for single-output settings and have no built-in mechanism to aggregate uncertainty across a multi-round, multi-agent trajectory. Precisely exploiting this trajectory structure is the core contribution of \methodname.

For consistency-based baselines (SE, SentSAR), naively extending them to every intermediate generation is prohibitively expensive: with $3$ agents, $3$ rounds, and $5$ samples per node, a single query already requires $3 \times 3 \times 5 = 45$ generations. For single-sample baselines (MSP, PPL, PE), aggregating over all intermediate generations is exactly what \methodfix~does (Tables~\ref{tab:ue-auroc} and~\ref{tab:ue_different_baselines}); \methodname~further improves on \methodfix~by replacing the uniform aggregation with credibility-aware updates.

To stress-test this design, we nevertheless ran SE, SE$^*$, SentSAR, and SentSAR$^*$ with their uncertainty averaged across every intermediate and final generation in the trajectory. Table~\ref{tab:baselines_mean} reports the result on Llama 4 Scout under Debate. Even with this much richer signal, the consistency-based baselines remain below \methodname~on every dataset, with an average gap of $5.15$ points over the strongest contender.

\begin{table}[!h]
\centering
\caption{AUROC (\%) on Llama 4 Scout under Debate when consistency-based baselines are aggregated over every intermediate and final generation (``Mean over trajectory''), compared with \methodname. Per row: \textbf{best} / \underline{second-best}.}
\label{tab:baselines_mean}
\begin{sc}
\setlength{\tabcolsep}{8pt}
\renewcommand{\arraystretch}{1.15}
\begin{tabular}{@{}lccccc@{}}
\toprule
& \multicolumn{4}{c}{\textbf{Mean over trajectory}} & \\
\cmidrule(lr){2-5}
\textbf{Dataset} & \textbf{SE} & \textbf{SE$^{*}$} & \textbf{SentSAR} & \textbf{SentSAR$^{*}$} & \textbf{\methodname} \\
\midrule
MMLUP    & 54.30 & 65.48 & 65.50 & \underline{80.31} & \textbf{80.90} \\
AQUA     & 56.44 & 69.60 & \underline{76.36} & 74.97 & \textbf{79.77} \\
MedMC    & 51.31 & 52.99 & 61.16 & \underline{64.19} & \textbf{74.38} \\
MBPP     & 42.92 & \underline{64.09} & 59.91 & 59.52 & \textbf{64.56} \\
\midrule
\textit{Ave} & 51.49 & 64.54 & 65.73 & \underline{69.75} & \textbf{74.90} \\
\bottomrule
\end{tabular}
\end{sc}
\end{table}

\paragraph{All recorded methods over all settings.} We further extend this comparison to every UQ method whose per-node values are recorded, over all 20 settings of Table~\ref{tab:ue-auroc}. Each method's final-response score is replaced by the unweighted mean of its per-node values over the trajectory, i.e., the same aggregation rule as \methodfix. As shown in Table~\ref{tab:baselines_mean_all}, trajectory-level aggregation improves every method. Among them, the cost-free log-probability signal (i.e., \methodfix) yields the strongest aggregated score, and \methodname~improves on it further.

\begin{table}[!h]
\centering
\caption{Mean AUROC (\%) over the 20 settings of Table~\ref{tab:ue-auroc} when each method is computed on the final response (``Final node'') or averaged over the trajectory with the same rule as \methodfix~(``Trajectory'').}
\label{tab:baselines_mean_all}
\begin{sc}
\setlength{\tabcolsep}{8pt}
\renewcommand{\arraystretch}{1.15}
\begin{tabular}{@{}lccc@{}}
\toprule
\textbf{Method} & \textbf{Final node} & \textbf{Trajectory} & \textbf{Gain} \\
\midrule
MSP (trajectory $=$ \methodfix) & 62.09 & 70.91 & $+8.82$ \\
PPL         & 55.96 & 67.74 & $+11.78$ \\
PE          & 61.65 & 65.17 & $+3.52$ \\
SentSAR$^*$ & 61.16 & 64.59 & $+3.43$ \\
SentSAR     & 58.95 & 61.02 & $+2.07$ \\
SE$^*$      & 54.86 & 59.12 & $+4.26$ \\
SE          & 49.37 & 50.32 & $+0.95$ \\
\midrule
\methodname & --- & \textbf{74.47} & --- \\
\bottomrule
\end{tabular}
\end{sc}
\end{table}

\subsection{Expected Calibration Error}\label{app:ece}

While AUROC measures the ranking quality of uncertainty estimates---whether higher uncertainty correlates with incorrect responses---it does not directly assess calibration. We additionally evaluate Expected Calibration Error (ECE; lower is better) on Llama 4 Scout under Debate. As shown in Table~\ref{tab:ece}, \methodname~achieves the best average ECE, confirming that improvements extend beyond ranking to calibration.

\begin{table}[!h]
\centering
\caption{Expected Calibration Error (\%, lower is better) on Llama 4 Scout under Debate. Per row: \textbf{best} / \underline{second-best}.}
\label{tab:ece}
\begin{sc}
\footnotesize
\setlength{\tabcolsep}{2.5pt}
\resizebox{\linewidth}{!}{%
\begin{tabular}{@{}lcccccccccccccc@{}}
\toprule
\textbf{Dataset} & \textbf{PPL} & \textbf{MSP} & \textbf{SentSAR} & \textbf{PE} & \textbf{SE} & \textbf{SE$^*$} & \textbf{SentSAR$^*$} & \textbf{CoCoA} & \textbf{CoCoA$^*$} & \textbf{Deg$^*$} & \textbf{Deg} & \textbf{G-NLL} & \textbf{\methodfix} & \textbf{\methodname} \\
\midrule
MMLUP    & \underline{9.77} & 11.70          & 28.94 & 10.15          & 27.50 & 20.32 & 18.18 & 12.68 & 18.90 & 12.17            & 12.79 & 12.71            & 10.59             & \textbf{6.25} \\
AQUA     & 8.89             & \textbf{2.98}  & 40.60 & 11.68          & 22.08 & 10.07 & 8.70  & 3.76  & 9.58  & 12.05            & 20.80 & \underline{3.70} & 4.50              & 5.50 \\
MedMC    & 19.47            & 18.86          & 24.84 & 16.32          & 28.31 & 22.91 & 19.28 & 20.47 & 24.20 & \underline{8.75} & 15.55 & 18.17            & 10.03             & \textbf{7.63} \\
MBPP     & 35.73            & 36.11          & 19.04 & 31.79          & 40.48 & 27.94 & 19.60 & 36.32 & 36.54 & 29.71            & 29.08 & 34.81            & \underline{18.40} & \textbf{15.78} \\
\midrule
\textit{Ave} & 18.46        & 17.41          & 28.36 & 17.48          & 29.59 & 20.31 & 16.44 & 18.31 & 22.30 & 15.67            & 19.55 & 17.35            & \underline{10.88} & \textbf{8.79} \\
\bottomrule
\end{tabular}%
}
\end{sc}
\end{table}

\subsection{Normalization Functions}\label{app:normalize_function}

\begin{table}
\centering
\caption{AUROC (\%) results. ``Ave'' represents the mean over the eight dataset cells in each row.}
\label{tab:normalize}
\begin{sc}
\setlength{\tabcolsep}{4pt}
\resizebox{\linewidth}{!}{%
\begin{tabular}{@{}lccccccccc@{}}
\toprule
\multicolumn{10}{c}{\textbf{Llama 4 Scout}} \\
\midrule
& \multicolumn{4}{c}{\textbf{Debate}} & \multicolumn{4}{c}{\textbf{DyLAN}} & \\
\cmidrule(lr){2-5} \cmidrule(lr){6-9}
\textbf{Method} & \textbf{MMLUP} & \textbf{AQUA} & \textbf{MedMC} & \textbf{MBPP} &
\textbf{MMLUP} & \textbf{AQUA} & \textbf{MedMC} & \textbf{MBPP} & \textbf{Ave} \\
\midrule
power            & 80.43 & 79.99 & 74.12 & 64.22 & 83.82 & 80.84 & 75.54 & 53.80 & 74.10 \\
exponential      & 81.36 & 79.89 & 73.74 & 63.82 & 82.95 & 80.06 & 75.43 & 54.93 & 74.02 \\
softmax          & 81.89 & 79.64 & 74.66 & 64.84 & 84.51 & 80.44 & 73.74 & 53.24 & 74.12 \\
min-max          & 81.21 & 79.39 & 74.62 & 65.05 & 83.53 & 81.41 & 74.09 & 53.47 & 74.10 \\
tanh             & 80.74 & 79.97 & 74.15 & 64.00 & 83.71 & 80.71 & 75.97 & 54.00 & 74.16 \\
sigmoid          & 80.90 & 79.77 & 74.38 & 64.56 & 83.70 & 80.63 & 76.16 & 56.98 & 74.64 \\
\midrule\midrule
\multicolumn{10}{c}{\textbf{GPT-4o mini}} \\
\midrule
& \multicolumn{4}{c}{\textbf{Debate}} & \multicolumn{4}{c}{\textbf{DyLAN}} & \\
\cmidrule(lr){2-5} \cmidrule(lr){6-9}
\textbf{Method} & \textbf{MMLUP} & \textbf{AQUA} & \textbf{MedMC} & \textbf{MBPP} &
\textbf{MMLUP} & \textbf{AQUA} & \textbf{MedMC} & \textbf{MBPP} & \textbf{Ave} \\
\midrule
power           & 74.19 & 86.37 & 69.94 & 61.06 & 76.43 & 87.40 & 70.33 & 61.14 & 73.36 \\
exponential     & 74.32 & 85.98 & 71.36 & 58.62 & 74.33 & 86.55 & 66.26 & 61.27 & 72.34 \\
softmax         & 75.64 & 85.09 & 71.37 & 59.94 & 76.54 & 85.32 & 71.47 & 60.11 & 73.19 \\
min-max         & 75.39 & 85.27 & 71.55 & 59.30 & 76.99 & 88.07 & 68.60 & 59.07 & 73.03 \\
tanh            & 73.94 & 85.62 & 70.76 & 60.78 & 76.75 & 87.50 & 70.82 & 61.63 & 73.47 \\
sigmoid         & 74.40 & 85.51 & 71.03 & 60.37 & 76.54 & 87.48 & 71.12 & 61.69 & 73.52 \\
\bottomrule
\end{tabular}%
}
\end{sc}
\vspace{-0.5em}
\end{table}

In this section, we compare different normalization functions in Eq.~\ref{eq:normalize_weight} to map the raw edge-level credibility score
$w_{j\to i}^{(t-1)}$ into a bounded update weight in $[0,1]$.
The resulting normalized weight is used in the node-level aggregation step
and subsequently modulates uncertainty updates.
For clarity, we denote the normalized value by $\hat{w}_{j\to i}^{(t-1)}\in[0,1]$.

\paragraph{(1) \textbf{Tanh normalization.}}
We apply a smooth, sign-preserving squashing function and map to $[0,1]$:
\begin{equation}
    \hat{w}_{j\to i}^{(t-1)} = \frac{1}{2}\tanh\!\big(w_{j\to i}^{(t-1)}\big) + \frac{1}{2}.
    \label{eq:norm_tanh}
\end{equation}

\paragraph{(2) \textbf{Sigmoid normalization.}}
We use the logistic function to map $w_{j\to i}^{(t-1)}$ to $[0,1]$:
\begin{equation}
    \hat{w}_{j\to i}^{(t-1)} = \sigma\!\big(w_{j\to i}^{(t-1)}\big)
    = \frac{1}{1+\exp\!\big(-w_{j\to i}^{(t-1)}\big)}.
    \label{eq:norm_sigmoid}
\end{equation}

\paragraph{(3) \textbf{Min--max normalization.}}
For a fixed update node $v_i^{(t)}$, let $\sR_i^{(t)}$ denote its peer set and consider the collection
$\{w_{k\to i}^{(t-1)}\}_{k\in\sR_i^{(t)}}$. We rescale within this peer set:
\begin{equation}
    \hat{w}_{j\to i}^{(t-1)} =
    \begin{cases}
    \dfrac{w_{j\to i}^{(t-1)}-\min_{k\in\sR_i^{(t)}} w_{k\to i}^{(t-1)}}
    {\max_{k\in\sR_i^{(t)}} w_{k\to i}^{(t-1)}-\min_{k\in\sR_i^{(t)}} w_{k\to i}^{(t-1)}},
    & \text{if } \max-\min > \varepsilon,\\[8pt]
    \dfrac{1}{2}, & \text{otherwise,}
    \end{cases}
    \label{eq:norm_minmax}
\end{equation}
where $\varepsilon$ is a small constant to avoid numerical instability when the range collapses.

\paragraph{(4) \textbf{Softmax normalization (peer-set dependent).}}
We convert the peer-wise edge scores into a distribution over peers:
\begin{equation}
    \hat{w}_{j\to i}^{(t-1)} =
    \frac{\exp\!\big(w_{j\to i}^{(t-1)}\big)}
    {\sum_{k\in\sR_i^{(t)}} \exp\!\big(w_{k\to i}^{(t-1)}\big)}.
    \label{eq:norm_softmax}
\end{equation}
This emphasizes relative ranking among peers.

\paragraph{(5) \textbf{Exponential mapping (magnitude-based, asymmetric).}}
We map the magnitude of $w_{j\to i}^{(t-1)}$ to $[0,1]$ via an exponential transform and handle negative
scores asymmetrically:
\begin{equation}
    \hat{w}_{j\to i}^{(t-1)} =
    \begin{cases}
    1-\exp\!\big(-|w_{j\to i}^{(t-1)}|\big), & w_{j\to i}^{(t-1)}\ge 0,\\
    \exp\!\big(-|w_{j\to i}^{(t-1)}|\big),   & w_{j\to i}^{(t-1)}<0.
    \end{cases}
    \label{eq:norm_exponential}
\end{equation}

\paragraph{(6) \textbf{Power mapping.}}
We apply a sign-preserving power transform and map to $[0,1]$:
\begin{equation}
    \hat{w}_{j\to i}^{(t-1)} =
    \frac{1}{2}\Big(\mathrm{sign}\!\big(w_{j\to i}^{(t-1)}\big)\cdot|w_{j\to i}^{(t-1)}|^{\alpha}\Big)
    + \frac{1}{2},
    \label{eq:norm_power}
\end{equation}
where we use $\alpha=\tfrac{1}{2}$ (square-root) in our implementation.
\\
The results for different normalization functions are summarized in Table~\ref{tab:normalize}.

\begin{figure}[!htp]
    \centering
    \begin{subfigure}[b]{1\textwidth}
        \centering
        \includegraphics[width=\linewidth]{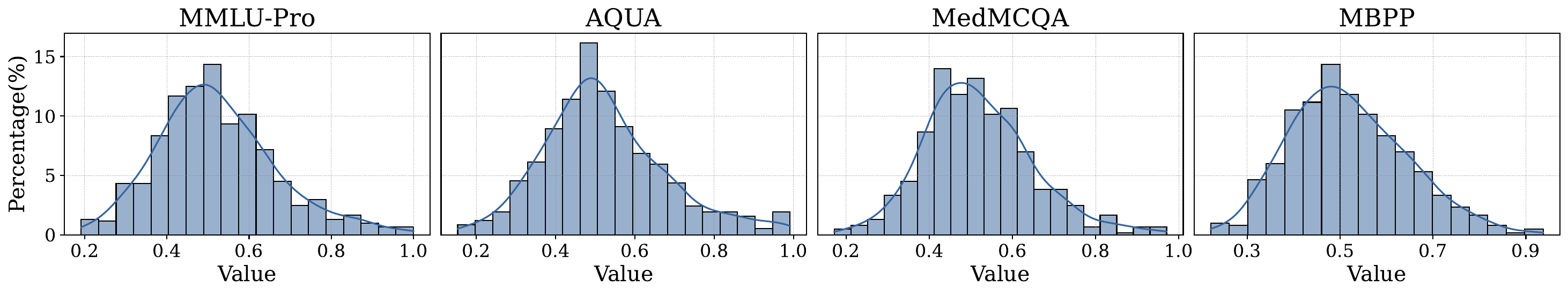}
        \caption{Llama 4 Scout}
        \label{fig:top}
    \end{subfigure}

    \begin{subfigure}[b]{1\textwidth}
        \centering
        \includegraphics[width=\linewidth]{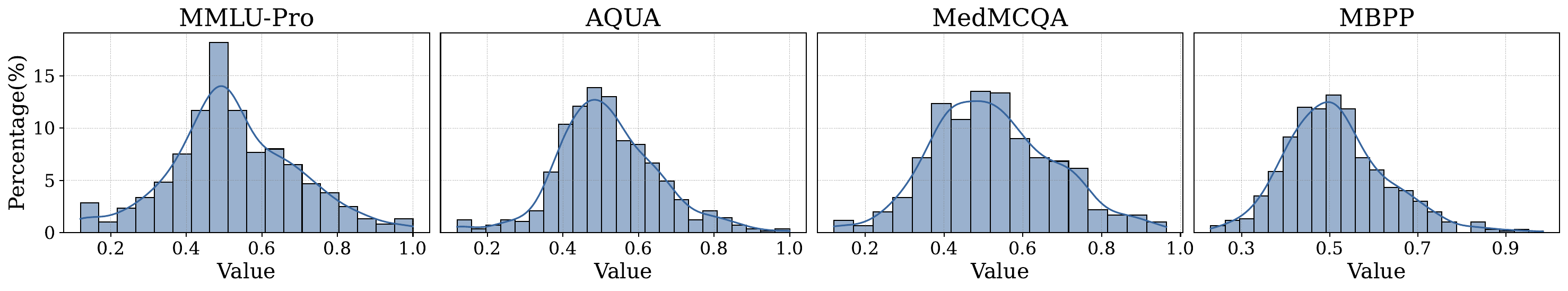}
        \caption{GPT-4o mini}
        \label{fig:bottom}
    \end{subfigure}

    \caption{The distribution of $w_i^{(t)}$ in Debate System}
    \label{fig:w_distribution}
\end{figure}

\subsection{Alternative Aggregation Strategies}\label{app:aggregation}

The system-level aggregation in Eq.~\ref{eq:umas} averages node-level uncertainty uniformly across the trajectory. To verify that the gains of \methodname~are robust to this design choice, we evaluate four alternatives:

\begin{itemize}
    \item \textbf{Depth-weighted}: later interaction rounds receive proportionally larger weight.
    \item \textbf{Exponential-weighted}: later rounds receive exponentially larger weight via $\exp(\text{round})$.
    \item \textbf{Min}: aggregate via the most-confident node only.
    \item \textbf{Max}: aggregate via the least-confident node only.
\end{itemize}

Results on GPT-4o mini under Debate are reported in Table~\ref{tab:aggregation}. All \methodname~variants substantially outperform every baseline in the main result table (Table~\ref{tab:ue-auroc}) regardless of aggregation strategy, confirming that the gains stem from trajectory-level aggregation and credibility-aware updating rather than the specific choice of aggregation function.

\begin{table}[!h]
\centering
\caption{AUROC (\%) on GPT-4o mini under Debate for alternative \methodname~aggregation strategies. Per row: \textbf{best} / \underline{second-best}.}
\label{tab:aggregation}
\begin{sc}
\setlength{\tabcolsep}{6pt}
\renewcommand{\arraystretch}{1.15}
\begin{tabular}{@{}lcccccc@{}}
\toprule
\textbf{Dataset} & \textbf{\methodfix} & \textbf{\methodname} & \textbf{\methodname-depth} & \textbf{\methodname-exp} & \textbf{\methodname-min} & \textbf{\methodname-max} \\
\midrule
MMLUP    & \textbf{77.42} & 74.40             & 75.06          & \underline{75.14} & 73.60 & 74.62 \\
AQUA     & 81.70          & \textbf{85.51}    & 84.30          & 83.20             & 84.05 & \underline{84.98} \\
MedMC    & 66.03          & \underline{71.03} & \textbf{71.06} & 70.88             & 70.62 & 69.94 \\
MBPP     & 57.62          & \underline{60.37} & 60.13          & 59.73             & 59.41 & \textbf{61.13} \\
\midrule
\textit{Ave} & 70.69          & \textbf{72.83}    & 72.64          & 72.24             & 71.92 & \underline{72.77} \\
\bottomrule
\end{tabular}
\end{sc}
\end{table}

\paragraph{Final-round-only scoring.} Averaging over all nodes may overestimate uncertainty for answers that were wrong in early rounds but corrected later in the interaction. Scoring only the final round removes this effect entirely, yet it is the weakest of the six aggregation rules we tested (the five above plus final-round-only): averaged over the 20 settings of Table~\ref{tab:ue-auroc}, it reaches 72.74 AUROC versus 74.47 for the uniform average. A trajectory that needed correction is riskier than one that was right from the start, so early nodes carry useful signal.

\subsection{Distribution of Weight}\label{app:weight_distribution}

In this section, we report statistics of the distribution of $w_i^{(t)}$ in Figure~\ref{fig:w_distribution} to demonstrate that it is a genuinely dynamic quantity that varies across inputs, rather than collapsing to a constant. We further observe that $w_i^{(t)}$ is approximately normally distributed, suggesting that peer credibility signals are typically moderate and fluctuate around a stable mean, with extremely high/low-credibility updates occurring relatively rarely.

\subsection{Robustness to the Correctness Definition}\label{app:extraction}

Exact-match evaluation depends on how the final answer is extracted from a free-form response, and extraction errors introduce label noise into the AUROC evaluation. We therefore re-label every MAS answer under alternative extraction strategies and recompute all AUROCs. Besides our default extractor (E1, Qwen3-Next-80B), we consider a regular-expression extractor (E2), a cascaded extractor (E3), an option-text matcher (E4), and re-extraction by a second LLM from a different model family (E5, Mistral-Small-3.2-24B), as well as two aggregates: \emph{strict} counts an answer as correct only if all applicable strategies agree, and \emph{lenient} if any does. E1 and E5 agree on 97.8\% of extracted answers and 99.5\% of correctness labels. Because MBPP is scored by pass@1 execution of the generated program and has no free-form answer span to extract, this analysis covers the four short-answer datasets, i.e., 16 of the 20 settings in Table~\ref{tab:ue-auroc}; E1 is reported on the same 16 settings.

Table~\ref{tab:extraction} shows that the ranking does not depend on the extraction strategy: \methodname~ranks first and \methodfix~second under all seven correctness definitions. Label noise shifts all methods together, and the margin of \methodname~over the best baseline never falls below $+8.19$ AUROC (E4).

\begin{table}[!h]
\centering
\caption{Mean AUROC (\%) over the 16 short-answer settings under alternative correctness definitions (see text). Per column: \textbf{best} / \underline{second-best}.}
\label{tab:extraction}
\begin{sc}
\setlength{\tabcolsep}{5pt}
\renewcommand{\arraystretch}{1.15}
\begin{tabular}{@{}lccccccc@{}}
\toprule
\textbf{Method} & \textbf{E1} & \textbf{E2} & \textbf{E3} & \textbf{E4} & \textbf{E5} & \textbf{Strict} & \textbf{Lenient} \\
\midrule
Best baseline (CoCoA$^*$) & 64.42 & 59.20 & 63.64 & 61.83 & 64.94 & 61.05 & 63.21 \\
\methodfix  & \underline{74.20} & \underline{66.46} & \underline{72.52} & \underline{69.32} & \underline{74.86} & \underline{68.55} & \underline{71.04} \\
\methodname & \textbf{77.87} & \textbf{69.46} & \textbf{75.65} & \textbf{70.02} & \textbf{78.64} & \textbf{71.12} & \textbf{73.98} \\
\bottomrule
\end{tabular}
\end{sc}
\end{table}

\subsection{Controlling for Response Length}\label{app:length}

MSP is the total sequence log-probability and therefore decreases with response length, while MAS generations vary considerably in length. To test whether \methodname~merely tracks verbosity, we run the following controls over the 20 settings of Table~\ref{tab:ue-auroc}, where length is the token count of the scored text. Results are summarized in Table~\ref{tab:length}.

\paragraph{Length-stratified AUROC.} Within each setting, questions are binned into length quartiles; AUROC is computed inside each bin and pooled by bin size. All methods except PPL lose AUROC under this control, but the ranking is unchanged: \methodname~retains 69.26, and its margin over the strongest baseline narrows from 11.95 to 8.20 points.

\paragraph{Incremental value over length.} $\Delta$AUC is the AUROC gained by adding a score to a logistic regression of correctness on length. \methodname~adds the most ($+5.56$, significant in 14 of 20 settings). We further fit a logistic regression of correctness on standardized MSP together with terminal and trajectory length; adding \methodname~to this model still gains $+4.05$ AUROC (significant in 14 of 20 settings), so \methodname~is not MSP with a length adjustment.

\paragraph{Length-normalized confidence.} PPL in Table~\ref{tab:ue-auroc} is the length-normalized counterpart of MSP. Rebuilding \methodname~on length-normalized (per-token) log-probabilities lowers its AUROC from 74.47 to 69.96 (\methodfix: from 70.91 to 67.74), so most of the gain survives. We keep the unnormalized log-probability, consistent with the finding of Aichberger et al.~\cite{aichberger2024rethinking} that length normalization degrades likelihood-based UQ even under highly variable response lengths.

\begin{table}[!h]
\centering
\caption{Length controls, averaged over the 20 settings of Table~\ref{tab:ue-auroc}. Raw: standard AUROC (\%). Sig.: number of settings in which $\Delta$AUC is significant.}
\label{tab:length}
\begin{sc}
\setlength{\tabcolsep}{8pt}
\renewcommand{\arraystretch}{1.15}
\begin{tabular}{@{}lcccc@{}}
\toprule
\textbf{Method} & \textbf{Raw} & \textbf{Within strata} & \textbf{$\Delta$AUC over length} & \textbf{Sig.} \\
\midrule
MSP         & 62.09 & 58.42 & $+1.84$ & 5/20 \\
PPL         & 55.96 & 57.91 & $+1.89$ & 5/20 \\
PE          & 61.65 & 61.06 & $+4.10$ & 12/20 \\
CoCoA$^*$   & 62.52 & 58.72 & $+2.03$ & 10/20 \\
\methodfix  & 70.91 & 66.94 & $+4.41$ & 11/20 \\
\methodname & \textbf{74.47} & \textbf{69.26} & $\mathbf{+5.56}$ & \textbf{14/20} \\
\midrule
Length only & 66.28 & 56.04 & --- & --- \\
\bottomrule
\end{tabular}
\end{sc}
\end{table}

\end{document}